\documentclass[
 reprint,
 superscriptaddress,
 amsmath,amssymb,
 aps,
 prx,
 longbibliography,
 floatfix
]{revtex4-2}
\usepackage{physics}
\usepackage{graphicx}
\usepackage{dcolumn}
\usepackage{bm}
\usepackage{xcolor}
\usepackage{hyperref}
\usepackage{tikz}
\usepackage[capitalise]{cleveref}
\usepackage{simpler-wick}
\usepackage{orcidlink}
\usetikzlibrary{matrix,fit,positioning}
\hypersetup{
  colorlinks=true,
  linkcolor=blue,
  citecolor=blue,
  urlcolor=blue
}
\usepackage{mathtools}
\renewcommand{\arraystretch}{1.3}
\definecolor{dqmcPurple}{HTML}{8E24AA}

\begin{document}

\preprint{APS/123-QED}

\title{Delocalized Coupled-Cluster Theory for Polaron Structure and Dynamics}

\author{Hamlin Wu \orcidlink{0009-0003-3690-2938}}
\author{Moritz K. A. Baumgarten \orcidlink{0000-0002-5214-2655}}
\author{Tong Jiang \orcidlink{0000-0002-5907-4886}}
\author{Joonho Lee \orcidlink{0000-0002-9667-1081}}

\affiliation{
Department of Chemistry and Chemical Biology, \\ Harvard University, Cambridge, Massachusetts, 02138, United States
}

\date{\today}
\begin{abstract}
Polaron ground states and finite-temperature dynamics remain challenging to simulate because existing methods struggle to combine nonperturbative accuracy, systematic improvability, and scalability from models to materials-specific Hamiltonians.
We introduce a translationally invariant variational coupled-cluster (CC) theory for polarons, termed delocalized CC (dCC), with closed-form energies at cost as low as $\mathcal{O}(N^3)$ and no phonon-number cutoff.
dCC accurately describes the ground states of the one- and two-dimensional Holstein and Su--Schrieffer--Heeger (optical and bond) models and the Fr{\"o}hlich model, in close agreement with density matrix renormalization group (DMRG) and diagrammatic Monte Carlo benchmarks.
A projected tangent-space response formalism built on the same ansatz yields electron-addition spectral functions and optical conductivities at zero and finite temperature.
The resulting spectra agree well with DMRG, Lanczos, and neural-network quantum-state benchmarks while retaining a physically interpretable excitation hierarchy, and extend to two-dimensional lattices at finite temperature beyond the practical reach of these methods.
The same framework applies directly to \textit{ab initio} electron--phonon matrix elements, yielding LiF electron- and hole-polaron binding energies that match state-of-the-art many-body calculations.
These results establish dCC as a unified variational framework for polaron ground states and dynamics, from model systems to real materials.
\end{abstract}
\maketitle
\section{Introduction}
A charge carrier dressed by the lattice distortion forms a polaron, the classic quasiparticle of electron--phonon physics \cite{Landau1933, pekar1946, Frohlich1954, holstein1959p1, holstein1959p2, Alexandrov2007, EminBook}.
Closely related fermion--boson dressing problems arise in light--matter coupled systems \cite{Haugland2020, Mandal2023, Mordovina2020, flick2017,Ruggenthaler2018}, magnetic materials \cite{Kane1989, Martinez1991, SchmittRink1988, Dagotto1994, Grusdt2019, Koepsell2019, prembabu2025}, and ultracold gases \cite{Schmidt2015, grusdt2015, Massignan2014, Shchadilova2016, Jorgensen2016, Hu2016}, where photons, magnons, or collective atomic excitations play the role of phonons. Polaron concepts have likewise shaped the theory of electron transfer and solvent reorganization in chemical physics \cite{Marcus1965, Marcus1964, Chandler1984, Donley2014}. These make the polaron problem a paradigmatic setting for developing many-body techniques of broad relevance.
Canonical model Hamiltonians capture distinct limits of electron--lattice coupling, including the long-range Fr\"ohlich model \cite{Frohlich1954} and the short-range Holstein~\cite{holstein1959p1, holstein1959p2} and Peierls models \cite{zhangssh, kagan1976, Carbone2021}, while modern density-functional perturbation theory~\cite{BaroniDFPT, EPW, GiustinoRMP, QE2017} now enables such Hamiltonians to be constructed for real materials with first-principles electron--phonon matrix elements, and experimental probes resolve quasiparticle formation, transport, and relaxation with increasing spatial, momentum, and temporal resolution \cite{Damascelli2003, Sobota2021}. 
Quantitatively accurate, material-specific polaron calculations are thus increasingly needed and within reach \cite{RobinsonAIPolaron, LeeBernardi2021, Chang2025, Sio2019, Sio2019PRB, Luo2022, Bartolome2022, baumgartendd2, Luodqmc, BernardiSVD, GiustinoReview, Franchini2021,jiang2026first}.

However, no existing approach yet provides a general, systematically improvable, and scalable route to polaron ground states and finite-temperature dynamics across both model and materials-specific Hamiltonians.
Exact diagonalization, Lanczos, and density matrix renormalization group (DMRG) methods produce benchmark-quality results but remain exponentially hard to extend beyond low-dimensional models to materials-specific \textit{ab initio} Hamiltonians~\cite{Jeckelmann1998, Bonca1999, jansen2022, jansen2020FT}.
Diagrammatic Monte Carlo (DiagMC) can deliver numerically exact ground-state energies even for \textit{ab initio} Hamiltonians \cite{Mischenko2003, Luodqmc}, but requires high expansion orders at strong coupling and numerical analytic continuation for dynamical response.
Simple variational wavefunctions are computationally efficient but of uneven accuracy across coupling regimes and often lack a clear hierarchy for systematic improvement \cite{Toyozawa1961, Zhao2023, Sun2013,baumgartendd2}, while neural-network quantum-state (NNQS) wavefunctions can be accurate but are limited by the network sizes affordable at realistic scales and by parameters that lack a direct physical interpretation~\cite{mahajan24}.
We close this gap by developing a variational coupled-cluster (CC) framework rooted in quantum chemistry~\cite{Coester1958, CoesterKummel1960, Cizek1966, BartlettMusial2007}.

CC theory includes interparticle correlations through a hierarchy of excitations ordered by physical relevance, each with different polynomial scaling, and has been applied across settings spanning nuclei, molecules, Bose fluids, and solid-state systems \cite{bishop1991, Gruneis2019, Gruber2018, McClain2017, Monkhorst1977, ArponenBose, LeidemannNuc, Hagen2014}. A recent application of conventional CC to the Holstein polaron established the feasibility of this route, though with limited accuracy~\cite{Yang2024}. In contrast to nearly all prior CC methods, we introduce an efficient, {\it variational} CC theory for polarons which can be readily applied from the model to the \textit{ab initio} setting, and which is unconventional in two principal ways.
First, because the problem has a single carrier, we obtain a closed-form expression for the variational energy, evading the standard non-variational projective treatment of CC. We derive this expression up to a second-order truncation of the excitation hierarchy ({\it vide infra}) and optimize the wavefunction directly, with the lowest-order theory scaling as $\mathcal{O}(N^3)$ with system size $N$. 
Because the bosonic sector is built on coherent states, the theory also requires no explicit phonon-number cutoff.
Second, we restore the crystal-momentum symmetry of the wavefunction via a projection operator, enabling direct calculation of polaron energy bands and momentum-resolved dynamics. Unlike conventional equation-of-motion CC (EOM-CC) approaches to band structures, which diagonalize a linear excitation manifold on a fixed CC ground state \cite{McClain2017}, each polaron band energy here is obtained by variationally optimizing the full momentum-projected ansatz within the corresponding crystal-momentum sector.
The approach belongs to the ``variation-after-projection'' family and is manifestly size-consistent, a property that existing variation-after-projection approaches in the literature generally lack \cite{Qiu2017}.
We refer to the resulting translationally invariant wavefunction hierarchy as the delocalized coupled-cluster (dCC) ansatz.

In addition, we describe a tangent-space response formalism, analogous in spirit to equation-of-motion \cite{Stanton1993, BartlettMusial2007, Krylov2008} and linear-response \cite{Koch1990, Monkhorst1977} coupled-cluster theories, which gives access to low-lying excited states, from which we compute electron-addition spectral functions and optical conductivities at zero and finite temperatures.
The same excitation hierarchy used to improve the ground-state wavefunction generates a systematically improvable response manifold~\cite{hirata2000high} that retains a direct physical interpretation in terms of electron, phonon, and coupled electron--phonon degrees of freedom.

This paper is organized as follows. Sections~\ref{sec:background}--\ref{sec:dynamics} develop the electron--phonon Hamiltonians, the dCC ansatz and its variational optimization, and the projected tangent-space response formalism. Sections~\ref{sec:ground}--\ref{sec:ab_initio} benchmark ground states for one- and two-dimensional Holstein and optical and bond Su--Schrieffer--Heeger (SSH) models and the Fr\"ohlich model, present zero- and finite-temperature spectral functions and optical conductivities, and compute LiF electron- and hole-polaron binding energies from the first-principles Hamiltonian, before we conclude in Section~\ref{sec:conclusion}.

\section{Background}\label{sec:background}
\subsection{Polaron Hamiltonians}\label{sec:polaron_hamiltonians}
Treating lattice vibrations within the harmonic approximation and expanding the electronic Hamiltonian to first order in the atomic displacements yields the standard linear-coupling electron--phonon Hamiltonian \cite{Mahan, BaroniDFPT, GiustinoRMP}.
We restrict our attention to this level of description, although higher-order couplings and phonon anharmonicity can be important in some systems.

We consider a general single-carrier electron--phonon Hamiltonian \cite{Mahan, Alexandrov2007, GiustinoRMP, BruusFlensberg, Bernardi2016} written in momentum space as $\hat H = \hat H_{\rm el}+\hat H_{\rm ph}+\hat H_{\rm ep}$, where the components take the form
\begin{equation}
\begin{aligned}
\hat H_{\rm el} &= \sum_{i_\mathbf{k}}\epsilon_{i_\mathbf{k}}a^\dagger_{i_\mathbf{k}}a_{i_\mathbf{k}}, \quad \hat H_{\rm ph} = \sum_{\nu_{\mathbf{q}}}\omega_{\nu_{\mathbf{q}}}b^\dagger_{\nu_\mathbf{q}}b_{\nu_\mathbf{q}} 
\\\hat H_{\rm ep} &= \sum_{\mathbf{k} \mathbf{q} ij \nu} g^\nu_{ij}(\mathbf{k},\mathbf{q}) a^\dagger_{i_{\mathbf{k}+\mathbf{q}}}a_{j_\mathbf{k}}\left(b^\dagger_{\nu_{-\mathbf{q}}}+b_{\nu_\mathbf{q}}\right).
\end{aligned}
\label{eq:mom_ham}
\end{equation}
Here, $a^\dagger_{i_\mathbf{k}}$ and $a_{i_\mathbf{k}}$ create and annihilate a carrier in electronic band $i$ with crystal momentum $\mathbf{k}$, while $b^\dagger_{\nu_\mathbf{q}}$ and $b_{\nu_\mathbf{q}}$ create and annihilate a phonon in branch $\nu$ with crystal momentum $\mathbf{q}$. The quantities $\epsilon_{i_\mathbf{k}}$, $\omega_{\nu_{\mathbf{q}}}$, and $g^\nu_{ij}(\mathbf{k},\mathbf{q})$ define the bare electronic dispersion, harmonic phonon spectrum, and electron--phonon coupling matrix elements, respectively. 

Different choices of $\epsilon_{i_\mathbf{k}}$, $\omega_{\nu_{\mathbf{q}}}$, and $g^\nu_{ij}(\mathbf{k},\mathbf{q})$ correspond to the various model Hamiltonians considered below. For the single-band lattice models, we take the nearest-neighbor tight-binding dispersion on a $d$-dimensional hypercubic lattice, $\epsilon_{\mathbf{k}}=-2t\sum_{\mu=1}^{d}\cos k_\mu$. The Holstein model \cite{holstein1959p1, holstein1959p2} is obtained from a single dispersionless optical phonon branch, $\omega_{\mathbf{q}}=\omega$, and a momentum-independent local coupling $g_{\rm H}(\mathbf{k},\mathbf{q})=g/\sqrt{N}$, where $N$ is the number of unit cells (equal here to the number of lattice sites); this corresponds to a carrier density coupling to a local lattice displacement, modulating the on-site electronic energy.
We also consider the optical \cite{SSH1979, MarchandPRL} and bond~\cite{zhangssh, Carbone2021, kagan1976} SSH (OSSH and BSSH) models, in which lattice displacements modulate the hopping amplitudes between neighboring sites rather than the local electronic energy. In $d$ spatial dimensions, both models contain $d$ dispersionless optical phonon branches, with the oscillators residing either on the sites or on the bonds, and couplings to branch $\mu$ given by $g^\mu_{\rm OSSH}(\mathbf{k},\mathbf{q})=\frac{2ig}{\sqrt{N}}\left[\sin(k_\mu+q_\mu)-\sin k_\mu\right]$ and $g^\mu_{\rm BSSH}(\mathbf{k},\mathbf{q})=\frac{g}{\sqrt{N}}\left[e^{i(k_\mu+q_\mu)}+e^{-ik_\mu}\right]$.

The canonical Fr\"ohlich model \cite{Frohlich1954, Mahan, Devreese2016} provides a complementary test of the dCC ansatz for long-range interactions within the linear electron--phonon models. It is the continuum specialization of \cref{eq:mom_ham} to a single parabolic band and a single dispersionless longitudinal-optical branch. In units where $\hbar = m= \omega_\text{LO} = 1$, one has $\epsilon_{\mathbf{k}} = |\mathbf{k}|^2/2$, $\omega_{\mathbf{q}} = 1$, and $g_{\text{F}}({\mathbf{q}}) =|\mathbf{q}|^{-1}\left({2^{3/2}\pi\alpha}/{\Omega}\right)^{1/2}$ \cite{Devreese2016}, where $\alpha$ is the dimensionless Fr\"ohlich coupling and $\Omega$ is the system volume.

In the \textit{ab initio} setting, the same Hamiltonian form in \cref{eq:mom_ham} is retained, but the quantities $\epsilon_{i_\mathbf{k}}$, $\omega_{\nu_{\mathbf{q}}}$, and $g^\nu_{ij}(\mathbf{k},\mathbf{q})$ are computed from first-principles density functional perturbation theory \cite{BaroniDFPT, GiustinoRMP, EPW} rather than specified analytically. This allows multiple electronic bands, multiple phonon branches, and realistic momentum-dependent electron--phonon couplings to be included without modifying the formal structure of the dCC ansatz. 

\subsection{Mean-field reference state}\label{sec:reference_state}
For a fixed electronic configuration, the linear electron--phonon coupling displaces the equilibrium of each harmonic oscillator, and the ground state of a displaced oscillator is a coherent state.
Within the harmonic approximation, the normal vibrational modes are independent, and the collective phonon state may thus be expressed as a product of coherent states generated by unitary displacement operators \cite{Heller2019, Heller2022},
\begin{equation}
    \ket{\boldsymbol{\alpha}} = \prod_{\nu_\mathbf{q}}\mathcal{D}_{\nu_\mathbf{q}}(\alpha_{\nu_\mathbf{q}})\ket{\text{vac}}_{\text{ph}}, \quad \mathcal{D}_{\nu_{\mathbf{q}}}(\alpha) = e^{\alpha b^\dagger_{\nu_{\mathbf{q}}} - \alpha^*b_{\nu_{\mathbf{q}}}}.
\end{equation}

We take as our zeroth-order reference a factorizable electron--phonon product state,
\begin{equation}
    \ket{\Psi_{\text{D}2}} = \sum_{j_{\mathbf{k}}}A_{j_{\mathbf{k}}}a^\dagger_{j_{\mathbf{k}}}\ket{\text{vac}}_{\text{el}} \otimes \ket{\boldsymbol{\alpha}},
\end{equation}
which is known in the model polaron literature as the Davydov D2 ansatz \cite{Zhao2023}. 
Here, the electronic coefficients $\{A_{j_\mathbf{k}}\}$ and the coherent-state amplitudes $\boldsymbol{\alpha} = \{\alpha_{\nu_\mathbf{q}}\}$ are the variational parameters. The same product structure underlies the \textit{ab initio} polaron equations obtained by Giustino and co-workers \cite{Sio2019, Sio2019PRB, GiustinoReview}. The D2 reference employed here may thus be regarded as the lattice-model analog of these \textit{ab initio} polaron wavefunctions~\cite{baumgartendd2}.

While the coherent-state structure efficiently captures lattice relaxation and large mean phonon occupations without any phonon-number cutoff, the D2 ansatz is unsatisfactory in at least two respects. First, the wavefunction is of a factorized form and therefore includes no explicit electron--phonon correlations. This deficiency is particularly pronounced in the intermediate coupling regime, where correlated fluctuations are especially important \cite{Alexandrov2007}. Second, the optimized D2 state generally corresponds to a broken-symmetry localized state and does not, in general, carry well-defined crystal momentum. Restoring translational symmetry has been shown to significantly improve the resulting ground-state description at a modest increase in computational cost~\cite {Zhao1997,Capek1987,Bolsterli1976,baumgartendd2}. 
\subsection{Conventional coupled-cluster state}\label{subsec:cc_rudiments}
To move beyond this factorizable manifold, we build a CC ansatz on top of the D2 reference,
\begin{equation}
    \ket{\Psi_{\text{CC}}} = e^{\hat{T}}\ket{\Psi_{\text{D}2}},
\end{equation}
where $\hat{T}$ contains electronic, phonon, and coupled electron--phonon excitations \cite{Yang2024,White2020, Haugland2020}. 
CC theory is a widely used exponential ansatz for correlated many-body wavefunctions, originally developed in nuclear physics and later established as one of the central pillars of modern electronic-structure theory \cite{Coester1958,CoesterKummel1960,BartlettMusial2007,CrawfordSchaefer2000, Cizek1966}. Its generality extends to coupled fermion--boson problems, in which the excitation operator may also include pure bosonic and coupled electron--boson excitations \cite{White2020, Haugland2020, Monkhorst1987}, with related expansions describing molecular vibrations \cite{Christiansen2004, Prasad}.

The excitation operators are most naturally expressed in a basis adapted to the reference. The optimized D2 state distributes the electron over the electronic bands, such that the band $j$ carries a weight $w_j$ and has a definite momentum distribution given by,
\begin{equation}
    w_j = \Big(\sum_{\mathbf{k}}\big| A_{j_\mathbf{k}}\big|^2\Big)^{1/2}, \qquad A^{(0)}_{j_\mathbf{k}} = \frac{A_{j_\mathbf{k}}}{w_j},
\end{equation}
from which we define the rotated operators $c^\dagger_{j_x} = \sum_{\mathbf{k}} A_{j_{\mathbf k}}^{(x)} a^\dagger_{j_{\mathbf k}}$, where for each band the coefficients $A^{(x)}_{j_\mathbf{k}}$ with $x\in[0,N_\mathbf{k}-1]$ form an arbitrary orthonormal completion of $A^{(0)}_{j_\mathbf{k}}$, so that the $c^\dagger_{j_x}$ span the full single-particle Hilbert space. In this basis the reference takes the compact form,
\begin{equation}
\ket{\Psi_{\text{D}2}} = \sum_j w_j c^\dagger_{j_0}\ket{\text{vac}}\otimes \ket{\boldsymbol{\alpha}},   
\end{equation}
with $\sum_j w_j^2 = 1$ for a single carrier. 

The coefficients appearing in $\hat{T}$ are the coupled-cluster amplitudes $\{\{t^{j_x}_{i_0}\}, \{t_{\nu_\mathbf{q}}\}, \{t^{j_x}_{i_0, \nu_\mathbf{q}}\}\dots\}$, which constitute the wavefunction parameters in CC calculations. At first order, we write $\hat T^{(1)} = \hat T_{\rm el}^{(1)} + \hat T_{\rm ph}^{(1)} + \hat T_{\rm ep}^{(1)}$ with
\begin{align}
    \hat T_{\rm el}^{(1)} &= {\sum_{ijx}}'t^{j_x}_{i_0} c^\dagger_{j_x} c_{i_0}, \\
    \hat T_{\rm ph}^{(1)} &=\sum_{\nu_{\mathbf q}} t_{\nu_{\mathbf q}} b^\dagger_{\nu_{\mathbf q}}, \\
    \hat T_{\rm ep}^{(1)} &= {\sum_{ijx\nu_{\mathbf q}}}'
    t^{j_x}_{i_0,\nu_{\mathbf q}}
    c^\dagger_{j_x} c_{i_0} b^\dagger_{\nu_{\mathbf q}},
    \label{eq:T1ep}
\end{align}
where the primed sums run over all bands $i$, target bands $j$, and target orbitals $x$, excluding the case $(j,x) = (i,0)$ where no excitation is made. 
At second order, we add pure two-phonon and coupled electron-two-phonon excitations to the cluster operator,
\begin{equation}
\label{eq:second_order}
\begin{aligned}
    \hat{T}^{(2)}_\text{ph} &= \frac{1}{2}\sum_{\substack{{\nu_1}_{\mathbf{q}_1}}{\nu_2}_{\mathbf{q}_2}}t_{{\nu_1}_{\mathbf{q}_1}{\nu_2}_{\mathbf{q}_2}}b^\dagger_{{\nu_1}_{\mathbf{q}_1}}b^\dagger_{{\nu_2}_{\mathbf{q}_2}}\\
    \hat{T}^{(2)}_{\text{ep}} &= \frac{1}{2}{\sum_{ijx {\nu_1}_{\mathbf{q}_1} {\nu_2}_{\mathbf{q}_2}}}'t^{j_x}_{i_0, {\nu_1}_{\mathbf{q}_1}{\nu_2}_{\mathbf{q}_2}}c^\dagger_{j_x}c_{i_0}b^\dagger_{{\nu_1}_{\mathbf{q}_1}}b^\dagger_{{\nu_2}_{\mathbf{q}_2}}.
\end{aligned}
\end{equation}
The full second-order truncation of the cluster operator is thus $\hat{T}^{(2)} = \hat{T}^{(1)} + \hat{T}^{(2)}_\text{ph} + \hat{T}^{(2)}_\text{ep}$.

We denote a truncation of the coupled-cluster hierarchy by CC-$X$-$SY$ following previous works \cite{Haugland2020,White2020}. Here, $X$ is the highest pure-phonon excitation order retained in $\hat{T}_\text{ph}$, and $Y$ lists the retained coupled electron--phonon excitation orders in $\hat{T}_\text{ep}$. The electronic sector is complete at first order for one carrier and is therefore omitted from the label. For example, CC-$1$-S1 retains only first-order excitations, while CC-$2$-S1, CC-$1$-S12, and CC-$2$-S12 add $\hat{T}^{(2)}_\text{ph}$, $\hat{T}^{(2)}_\text{ep}$, or both, respectively. The second-order phonon excitation is exactly that of the squeezing transformation \cite{GerryKnight}, and the second-order electron--phonon excitation couples a two-phonon excitation to charge-carrier hopping, a type of excitation rarely discussed in the polaron literature so far. The momentum-projected CC variants discussed below inherit the same notation with the dCC prefix.

Although $\hat{T}$ is truncated, the exponential structure of the ansatz generates products of the retained operators, producing an infinite sequence of higher-order disconnected excitations. In conventional CC, the cluster amplitudes $\{t_\mu\}$ are fixed projectively by requiring the similarity-transformed Schr\"odinger equation, $\bra{\Phi_\mu}e^{-\hat{T}}\hat{\mathcal{H}}e^{\hat{T}}\ket{\Phi} = 0$, to hold within the retained manifold, yielding nonlinear amplitude equations and an energy $E_\text{CC} = \bra{\Phi} e^{-\hat{T}}\hat{\mathcal{H}}e^{\hat{T}}\ket{\Phi}$ that is generally non-variational at finite truncation \cite{BartlettMusial2007, CrawfordSchaefer2000}.
In the context of single-polaron problems,
this projective strategy was implemented and tested in Ref. \citenum{Yang2024} with relatively inaccurate results.
As we show in \cref{sec:cc_polaron}, the single-carrier polaron problem instead admits a closed-form energy that can be minimized variationally with momentum projection. We determine the amplitudes this way throughout and achieve significantly greater accuracy than the projective results in Ref. \citenum{Yang2024} would suggest.

\section{Delocalized Coupled-Cluster Theory}\label{sec:cc_polaron}
We now specialize the framework of \cref{subsec:cc_rudiments} to lattice single-polaron problems, departing from conventional CC theory in two respects: crystal-momentum symmetry, which the reference breaks, is explicitly restored by projection, and the cluster amplitudes are determined variationally rather than projectively.

\subsection{Momentum projection and delocalized coupled-cluster ansatz}
One critical deficiency of $\ket{\Psi_{\text{CC}}}$ for single-polaron problems is that it does not, in general, respect the discrete translational symmetry of the lattice. 
As has been seen in many models and more recently in {\it ab initio} studies, such symmetry breaking
leads to significant errors when large polarons form \cite{baumgartendd2}.
One can restore this symmetry by projecting the broken-symmetry wavefunction onto a definite crystal momentum $\mathbf{K}$,
\begin{equation}
\label{eq:dCC}
    \ket{\Psi_{\text{dCC}}^{\mathbf{K}}} 
    = \hat{\Xi}_{\mathbf{K}} \ket{\Psi_{\text{CC}}}
    = \hat{\Xi}_{\mathbf{K}} e^{\hat{T}}\ket{\Psi_{\text{D}2}},
\end{equation}
with the momentum projector $\hat{\Xi}_{\mathbf{K}}$ given by
\begin{equation}
    \hat{\Xi}_{\mathbf{K}} = \frac{1}{\Omega}\sum_{\mathbf{R}}e^{i(\mathbf{K} - \hat{P})\cdot \mathbf{R}}
    =
    \frac1\Omega \sum_{\mathbf R}
    e^{i_\mathbf{K}\cdot \mathbf{R}}\hat{\mathcal{T}}_{\mathbf{R}},
\end{equation}
where $\mathbf{R}$ are real-space lattice vectors within the supercell, $\Omega$ their total number, $\hat{\mathcal T}_{\mathbf R}$ is the lattice translation operator, and $\hat{P} = \sum_{i_\mathbf{k}}\mathbf{k}\, a^\dagger_{i_\mathbf{k}}a_{i_\mathbf{k}} + \sum_{\nu_\mathbf{q}} \mathbf{q}\, b^\dagger_{\nu_\mathbf{q}}b_{\nu_\mathbf{q}}$ is the total crystal momentum operator. Under conjugation by $\hat{\mathcal{T}}_{\mathbf{R}}$, the creation operators acquire simple phase factors, $\hat{\mathcal{T}}_{\mathbf{R}} a^\dagger_{j_\mathbf{k}}\hat{\mathcal{T}}^{-1}_{\mathbf{R}} = e^{-i_\mathbf{k}\cdot \mathbf{R}}a^\dagger_{j_\mathbf{k}}$ and $\hat{\mathcal{T}}_{\mathbf{R}} b^\dagger_{\nu_{\mathbf{q}}}\hat{\mathcal{T}}_{\mathbf{R}}^{-1} = e^{-i\mathbf{q}\cdot \mathbf{R}}b^\dagger_{\nu_{\mathbf{q}}}$, as follows from a Baker--Campbell--Hausdorff expansion \cite{Sakurai_Napolitano_2020}.

In analogy with the delocalized constructions used in the Davydov hierarchy, we refer to the ansatz in \cref{eq:dCC} as the delocalized coupled-cluster (dCC) ansatz; the projected reference itself, $\hat{\Xi}_{\mathbf{K}}\ket{\Psi_{\text{D}2}}$, defines the delocalized D2 (dD2) ansatz \cite{Toyozawa1961, Zhao1997, baumgartendd2}. Although the resulting eigenstate carries a definite crystal momentum and is therefore translationally invariant, 
the polaron retains spatially localized two-point correlations as emphasized throughout the \textit{ab initio} polaron literature \cite{GiustinoReview, Bartolome2022}. This formulation is conceptually related to the Lee--Low--Pines (LLP) canonical transformation formulation \cite{Lee1953} in that it works within a fixed crystal-momentum sector. A key difference, however, is that here momentum conservation is enforced at the level of the wavefunction rather than by transforming the Hamiltonian into the center-of-mass frame.

\subsection{Variational energy evaluation of dCC}
Variational CC methods are known to scale exponentially with system size and are therefore generally impractical for many-electron systems, owing to the combinatorially growing number of terms generated by the combined action of excitations ($\hat{T}$) and de-excitations ($\hat{T}^\dagger$) on the many-electron reference \cite{Marie2021, CrawfordSchaefer2000, BartlettMusial2007}.
The single-polaron problem, however, is significantly simpler: because the fermionic sector contains only one electron, the electronic component of the coupled-cluster expansion naturally truncates at the one-electron excitation level. The bosonic sector remains infinite in principle, but is handled efficiently through the coherent-state structure of the reference. These simplifications allow us to obtain practical closed-form expressions for the variational polaron energy that can be evaluated at polynomial cost, in contrast to many-electron variational coupled-cluster theories \cite{Marie2021}, and to determine the CC amplitudes by direct variational energy minimization. Such a variational evaluation is more numerically stable than projective methods \cite{Yang2024}, and it avoids the approximations that arise in symmetry-restoring coupled-cluster theories evaluated in the usual projective fashion \cite{Qiu2017}.

For a total crystal momentum $\mathbf{K}$, the variational energy of dCC is given by
\begin{equation}
\label{eq:varE}
    E_{\text{dCC}}^{\mathbf{K}} = \frac{\bra{\Psi_{\text{dCC}}^{\mathbf{K}}}\hat{H} \ket{\Psi^{\mathbf{K}}_{\text{dCC}}}}{\bra{\Psi^{\mathbf{K}}_{\text{dCC}}}\ket{\Psi^{\mathbf{K}}_{\text{dCC}}}} = \frac{\sum_{\mathbf{R}}e^{i_\mathbf{K} \cdot \mathbf{R}}H(\mathbf{R})}{\sum_{\mathbf{R}}e^{i_\mathbf{K} \cdot \mathbf{R}}S(\mathbf{R})}.
\end{equation}
The second equality follows from the invariance of $\hat{H}$ under discrete lattice translations, $[\hat{H}, \hat{\Xi}_{\mathbf{K}}] = 0$, and the idempotency of the projector, $\hat{\Xi}_{\mathbf{K}}^2 = \hat{\Xi}_{\mathbf{K}}$: since the projector is a finite sum over real-space lattice vectors, the projected matrix elements reduce to overlaps and Hamiltonian matrix elements between the symmetry-broken CC wavefunction and its translated copies,
\begin{equation}
\begin{aligned}
   S(\mathbf{R}) &= \bra{\Psi_{\text{D}2}}e^{\hat{T}^\dagger}\hat{\mathcal{T}}_{\mathbf{R}} e^{\hat{T}}\ket{\Psi_{\text{D}2}},\\
    H(\mathbf{R}) &= \bra{\Psi_{\text{D}2}}e^{\hat{T}^\dagger}\hat{H} \hat{\mathcal{T}}_{\mathbf{R}}e^{\hat{T}}\ket{\Psi_{\text{D}2}}.
\end{aligned}
\end{equation}
One can obtain the dCC ground-state band structure by enumerating the crystal momenta $\mathbf{K}$ and minimizing the variational energy $E^{\mathbf{K}}_{\text{dCC}}$ with respect to the variational amplitudes.
This allows us to retain the full exponential structure of the CC ansatz at arbitrary $\mathbf{K}$, in contrast to standard CC, which would resort to equation-of-motion treatments that linearize the exponential.

The resulting expressions contain products of fermionic strings, bosonic strings, and translation-dependent phase factors. 
The fermionic part is significantly simplified by the one-electron structure, which causes exponentials carrying electronic excitations to truncate linearly without approximation. This allows us to write the dCC ansatz as 
\begin{equation}
\begin{aligned}
    \ket{\Psi^{\mathbf{K}}_{\text{dCC}}} &= \hat{\Xi}_{\mathbf{K}} e^{\hat{T}_{\text{el}}}e^{\hat{T}_{\text{ph}}}e^{\hat{T}_{\text{ep}}}\ket{\Psi_{\text{D}2}}\\ &= \hat{\Xi}_{\mathbf{K}}(1+\hat{T}_{\text{el}} + \hat{T}_{\text{ep}})e^{\hat{T}_{\text{ph}}}\ket{\Psi_{\text{D}2}}.
\end{aligned}
\end{equation}
The bosonic part can be evaluated analytically by exploiting the algebra of coherent states: the coherent phonon reference is an eigenstate of the phonon annihilation operator, $b_{ \nu_{\mathbf{q}}}\ket{\boldsymbol\alpha}=\alpha_{\nu_{\mathbf{q}}}\ket{\boldsymbol\alpha}$, and has overlap
\begin{equation}\label{eq:coherent_overlap}
\bra{\boldsymbol{\alpha}}\ket{\boldsymbol{\alpha}'} = e^{-\frac{1}{2}\sum_{\nu_{\mathbf{q}}} (|\alpha_{\nu_{\mathbf{q}}}|^2 + |\alpha'_{\nu_{\mathbf{q}}}|^2 - 2\alpha^{*}_{\nu_{\mathbf{q}}}\alpha_{\nu_{\mathbf{q}}}')}.
\end{equation}

\subsection{First-order dCC}

At first order, the dCC wavefunction separates naturally into two pieces,
\begin{equation}
\begin{aligned}
\label{eq:first_order_dcc}
    \ket{\Psi^{\mathbf{K}}_{\text{dCC}}} &= \hat{\Xi}_{\mathbf{K}}\left(\sum_{j\mathbf{k}}\zeta_{j,\mathbf{k}}a^\dagger_{j_\mathbf{k}}\ket{\mathrm{vac}}_{\text{el}}\otimes \ket{\boldsymbol{\Theta}}\right)\\
    &+\hat{\Xi}_{\mathbf{K}}\left(\sum_{j\mathbf{k}\nu_{\mathbf{q}}}\gamma^{j,\nu}_{\mathbf{k},\mathbf{q}}a^\dagger_{j_\mathbf{k}}\ket{\mathrm{vac}}_{\text{el}}\otimes b^\dagger_{ \nu_{\mathbf{q}}}\ket{\boldsymbol{\Theta}}\right).
\end{aligned}
\end{equation}
Here, the dressed coefficients $\zeta_{j,\mathbf{k}} = w_jA^{(0)}_{j_ \mathbf{k}}+{\sum_{ix}}'w_it^{j_x}_{i_0}  A^{(x)}_{j_\mathbf{k}}$ and $\gamma_{\mathbf{k},\mathbf{q}}^{j, \nu} = {\sum_{ix}}'w_it^{j_x}_{i_0, \nu_\mathbf{q}}A^{(x)}_{j_\mathbf{k}}$ collect the D2 parameters and CC amplitudes, and $\ket{\boldsymbol{\Theta}}$ denotes the coherent state resulting from the action of $e^{\hat{T}_{\text{ph}}}$ on the reference phonon state $\ket{\boldsymbol{\alpha}}$ (see Appendix~\ref{app:first_order}). The first piece corresponds to a dD2-component \cite{Zhao1997, Zhao2023, Toyozawa1961,baumgartendd2}, generated by the symmetry-restored D2 sector, while the second contains explicit electron--phonon correlations introduced through the coupled excitation operator $\hat{T}_{\text{ep}}$.

For the first-order theory, normal ordering of the bosonic operators appearing in the projected numerator and denominator reduces all bosonic matrix elements to polynomials in the coherent-state amplitudes multiplied by overlaps in \cref{eq:coherent_overlap}. In this way the infinite bosonic Hilbert space sums are resummed analytically into closed-form expressions that depend only on the variational parameters.

\subsection{Second-order dCC}
At second order, the structure of the variational manifold changes more substantially. In addition to the symmetry-restored coherent-state and one-phonon correlated sectors introduced at first order, the wavefunction contains two-phonon and coupled electron-two-phonon components generated by the corresponding second-order excitation operators. 
The pure phononic sector contains pair-creation terms (see \cref{eq:second_order}) whose exponential acting on the displaced coherent-state reference generates multimode squeezed coherent states \cite{GerryKnight}. Collectively, these terms encode pair-correlated phonon fluctuations and higher-order electron--phonon dressing processes beyond the first-order dCC manifold. While the associated matrix elements can no longer be reduced solely through the simple coherent-state eigenvalue relations used at first order, closed-form expressions for the energy remain tractable through the analytic evaluation of Gaussian integrals. We refer the reader to Appendix~\ref{app:secondorder} for the details of this derivation. 

\subsection{Scaling and implementation}
The dominant asymptotic scalings of the dCC hierarchy with respect to the number of lattice sites are summarized in \cref{tab:scaling}. At first order, the projected variational energy scales as $\mathcal{O}(N^3)$, while inclusion of either class of second-order excitations raises the scaling to $\mathcal{O}(N^4)$. Retaining the full second-order manifold leads to an overall scaling of $\mathcal{O}(N^5)$. We note that additional scaling reductions could be achieved by using low-rank factorized electron--phonon interaction matrix elements \cite{baumgartendd2}, which may require a different contraction strategy for evaluating the variational energy expressions. 

\begin{table}[t]
\caption{\label{tab:scaling}
Summary of the computational scaling of the dCC hierarchy for model systems with $N$ lattice sites. Scalings below do not factor in savings made possible through the use of low-rank factorizations of the electron--phonon interaction kernel.
}
\begin{ruledtabular}
\renewcommand{\arraystretch}{1.15}
\begin{tabular}{lcc}
Theory & Additional excitation introduced & Scaling \\
\colrule
dD2 & None & $\mathcal O(N^2)$ \\
dCC-1-S1   & Electron-one-phonon  & $\mathcal O(N^3)$ \\
dCC-1-S12  & Electron-two-phonon  & $\mathcal O(N^4)$ \\
dCC-2-S1   &Multimode squeezing & $\mathcal O(N^4)$ \\
dCC-2-S12  & Full second-order manifold & $\mathcal O(N^5)$ \\
\end{tabular}
\end{ruledtabular}
\end{table}

We implement the first-order variational expressions in momentum space, where translational structure can be exploited more directly. In this setting, FFT-based evaluation reduces the cost of selected intermediate quantities and improves the method's practical efficiency, while retaining the same asymptotic scaling as the lowest-order theory. In practice, the first-order dCC theory captures most of the relevant ground-state correlation across a broad range of regimes, while the second-order extensions, which we have implemented only in real space, provide further variational improvement in more challenging cases. We will showcase numerical examples of both first- and second-order dCC in \cref{sec:ground}.

\section{Dynamics from Tangent-Space Response}\label{sec:dynamics}
To access dynamical properties beyond the ground state, we construct a linear-response theory in the tangent space of the optimized dCC variational manifold. The central idea is that low-lying excited states can be well approximated as fluctuations about the variational minimum, generated by applying a linear set of excitation operators to the optimized ground state. 
These excitations are drawn from the same hierarchy used in the ground-state dCC theory, so the response formalism remains fully consistent with the underlying variational ansatz and inherits its systematic improvability.

In the following discussion, we take $\ket{\Psi^{\mathbf{K}}_{\text{dCC}}(\mathbf{t})}$ to be the optimized ground-state dCC wavefunction with a target crystal momentum $\mathbf{K}$, parameterized by variational amplitudes $\mathbf{t} = \{t_\mu\}$, with $\mathbf t_\text{ground}$ denoting their optimized ground-state values. For simplicity, we present the formalism assuming a single electronic and phonon band in real space, with the necessary generalization following from the discussion above.

\subsection{Tangent-space response manifold}
As translational symmetry is restored at the level of the wavefunction via the momentum projector, the corresponding response space is constructed to preserve it. We therefore parameterize the excited states $\ket{\Psi^{\mathbf{K}}_m}$, indexed by $m$, as 
\begin{equation}
    \ket{\Psi^{\mathbf{K}}_m} = \hat{\Xi}_{\mathbf{K}}\hat{\mathcal{R}}_m\ket{\Psi_{\text{CC}}(\mathbf{t})}
\end{equation}
where the excitation operator $\hat{\mathcal{R}}_m$ is expanded as 
\begin{equation}
\label{eq:excitation_EOM}
    \hat{\mathcal{R}}_m = \sum_{n \mu}\lambda_{\mu m}^{(n)} \hat{\mathcal{R}}_\mu^{(n)}  
\end{equation}
where $\mu$ labels the excitation type and $n$ its order, drawn from the same elementary excitation classes that appear in the dCC hierarchy.

Since $\ket{\Psi^{\mathbf{K}}_{\text{dCC}}(\mathbf{t})} = \hat{\Xi}_{\mathbf{K}}\ket{\Psi_{\text{CC}}(\mathbf{t})}$, these response states may also be interpreted as tangent vectors to the projected variational manifold, i.e., as first derivatives of the optimized wavefunction with respect to the variational amplitudes $\{t_\mu\}$ evaluated at $\mathbf{t} = \mathbf{t}_\text{ground}$ (see Appendix~\ref{app:tangent}).
This construction is closely related in spirit to equation-of-motion CC (EOM-CC) and linear-response CC theories, in which excited states are generated by acting on a correlated CC ground state with a hierarchy of excitation operators and are obtained by diagonalizing an effective Hamiltonian in the corresponding excitation manifold \cite{Stanton1993, Koch1990, BartlettMusial2007, Krylov2008, Monkhorst1977}. Here, however, the response manifold is constructed from momentum-projected excitation operators acting on the unprojected CC state. The resulting states form a nonorthogonal tangent-space basis, which is used to evaluate Hamiltonian and overlap matrix elements directly. This leads to a variational estimate of the low-lying excited-state spectrum in the spirit of more general tangent-space approaches for variational ans\"atze \cite{Shi2018, Hackl2020, marijanovic2026schrieffer}.

At first order, the elementary response operators are
\begin{equation}
    \hat{\mathcal{R}}^{(1)}_{\mu} \in \{c^\dagger_a c_i, b^\dagger_m, c^\dagger_a c_i b^\dagger_m\}
\end{equation}
which represent the leading-order electronic, phonon, and coupled electron--phonon fluctuations on top of the correlated dCC ground state, with corresponding basis states $\ket{\Psi^{\mathbf{K}}_\mu} = \hat{\Xi}_{\mathbf{K}}\hat{\mathcal{R}}^{(1)}_\mu\ket{\Psi_{\text{CC}}(\mathbf{t})}$.
To describe fluctuations beyond the first-order tangent space, we enlarge the response space by truncating \cref{eq:excitation_EOM} at higher orders, including the full set of linearly independent second-order sectors,
\begin{equation}
    \hat{\mathcal{R}}^{(2)}_\mu \in \{c^\dagger_a c_i b^\dagger_m b^\dagger_l , b^\dagger_m b^\dagger_l\},
\end{equation}
which correspond to coupled electron-two-phonon and pure two-phonon excitations. These sectors are essential for recovering response channels associated with redistributed spectral weight and multiphonon structure not captured at the first order.

Within the resulting basis at either order, we evaluate the projected Hamiltonian and overlap matrices,

\begin{equation}
    \tilde{H}_{\mu\nu}^{{\mathbf{K}}} = \bra{\Psi^{\mathbf{K}}_{\mu}}\hat{H} \ket{\Psi^{\mathbf{K}}_\nu}, \qquad S_{\mu\nu}^{{\mathbf{K}}} = \bra{\Psi^{\mathbf{K}}_{\mu}}\ket{\Psi^{\mathbf{K}}_{\nu}},
\end{equation}
and solve the generalized eigenvalue problem
\begin{equation}
\label{eq:generalev}
    \tilde{\mathbf{H}}^{\mathbf{K}}\mathbf{C} = \mathbf{S}^\mathbf{K} \mathbf C \boldsymbol{\varepsilon}.
\end{equation}
The eigenvectors obtained from \cref{eq:generalev} define the approximate excited states within the tangent manifold, $\ket{\Psi^{\mathbf{K}}_m} = \sum_{\mu}C_{\mu m} \ket{\Psi^{\mathbf{K}}_\mu}$, with associated energies $\varepsilon_m$.

While one could in principle construct a linear-response theory around a higher-order dCC ground state, in the present work we focus on dynamics built on top of the first-order optimized ground state (dCC-1-S1). This is similar in spirit to EOM-CC(n,m) in many-body electronic structure theory, proposed by Hirata and co-workers~\cite{hirata2000high}. To distinguish the excitation level used in the response space from that used in the ground state, we indicate the order of the linear-response manifold in parentheses, dCC-X-SY-$(n,m)$, where $n$ and $m$ are the highest retained orders in the phonon and coupled electron--phonon response sectors, respectively; as in the ground-state theory, the electronic sector terminates at first order and is not denoted.
For example, dCC-1-S1-(2,1) denotes a first-order dCC ground state combined with a response space containing up to two-phonon excitations and single coupled electron--phonon excitations.

Only the balanced manifolds dCC-X-SY-(n,n), where the phonon and coupled electron--phonon response sectors are expanded to the same order, define a systematic response hierarchy. Mixed manifolds with $n\neq m$, such as $(1,2)$ and $(2,1)$, selectively enlarge one response sector and are thus useful for diagnosing sector sensitivity, but possess imbalances which may be strongly amplified by thermally activated response channels at finite temperatures. Thus, systematic finite-temperature convergence should be assessed through the sequence of balanced manifolds.

In practice, the matrices are organized in blocks of electronic, phononic, coupled electron--phonon, two-phonon, and coupled electron-two-phonon sectors built on top of the dCC ground state, which is itself included as the zeroth-order vector of the resulting nonorthogonal response basis, with its metric accounted for through $\mathbf S^\mathbf{K}$; the block structure (\cref{fig:EOM}) and implementation details are given in Appendix~\ref{app:tangent}.

\subsection{Zero-temperature response from the tangent space}\label{subsec:response}
Once the projected tangent-space manifold has been constructed, zero-temperature dynamical observables follow by restricting the standard Lehmann and Kubo representations \cite{Mahan} to this variationally optimized spectrum. In the present work, we focus on the electron-addition spectral function and the optical conductivity.

\subsubsection{Electron-addition retarded Green's function}
For the single polaron problem, the natural single-particle quantity is the electron-addition retarded Green's function, $G_{\mathbf{k}}(t) = -\mathrm{i}\theta(t)\bra{0}a_{\mathbf{k}}(t) a_{\mathbf{k}}^\dagger(0)\ket{0}$, where $\ket{0}$ denotes the joint phonon and electron vacuum~\cite{Mahan}. Restricting its Lehmann expansion to the approximate excited states $\{\ket{\Psi^{\mathbf{K}}_{m}}\}$ and their eigenvalues $\{E_m^{\mathbf{K}}\}$ obtained from the generalized eigenvalue problem gives
\begin{equation}
\label{eq:freq_spec}
    G_{\mathbf{k}}(\omega) \simeq \sum_{\mathbf K}\sum_m \frac{|\bra{\Psi_{m}^{\mathbf{K}}}a^\dagger_{\mathbf{k}}\ket{0}|^2}{\omega - E_m^\mathbf{K} + \mathrm{i}\eta},
\end{equation}
where $\eta>0$ provides a Lorentzian broadening. Because the injected electron fixes the crystal momentum of the final state, only the sector with $\mathbf{K} = \mathbf{k}$ contributes. The spectral function follows from the imaginary part, $\mathcal{A}(\mathbf{k},\omega) = -\pi^{-1}\Im \,G_{\mathbf{k}}(\omega)$, and represents the electron-addition part of the single-particle spectrum most directly probed by inverse photoemission \cite{Damascelli2003, Sobota2021}.

\subsubsection{Optical conductivity}

We next consider the zero-temperature optical conductivity, which measures the linear current response of the polaron ground state to an applied electric field. In general, the ``regular'' part of this response can be obtained in the Lehmann representation from the Kubo formula~\cite{Mahan},
\begin{equation}
\label{eq:reg_oc}
\Re\,\sigma_{\alpha\alpha}^{\text{reg}}(\omega) =\frac{\pi}{\Omega}\sum_{m\neq 0}\frac{|\langle{\Psi_m^{\mathbf{K}_0}}|\,\hat{j}_{\alpha}|{\Psi_0}\rangle|^2}{E_m^{\mathbf{K}_0} - E_0}\delta(\omega - (E_{m}^{\mathbf{K}_0} - E_0)),
\end{equation}
where $\ket{\Psi_0}$ is the interacting ground state, $\mathbf{K}_0$ denotes the crystal momentum of the ground state, $\ket{\Psi_m^{\mathbf{K}_0}}$ are excited states within that sector, $\Omega$ is the system volume, and $\hat{j}_\alpha$ is the current operator along direction $\alpha$. Since the current operator is translationally invariant, it preserves crystal momentum and thus only couples states within the same momentum sector. For a given lattice Hamiltonian with a single electronic band, $\hat{j}_{\alpha}$ may be obtained from the continuity equation for the particle density \cite{Mahan}, from which it can be related to the time derivative of the polarization 
$\hat{\mathbf{P}} = e\sum_{\mathbf{n}}\mathbf{R}_{\mathbf{n}}a^\dagger_{\mathbf{n}}a_{\mathbf{n}}$ through the Heisenberg equations of motion as 
$
    \hat{\mathbf{j}} = {\partial \hat{\mathbf{P}}}/{\partial t} = \mathrm{i}[\hat{H}, \hat{\mathbf{P}}].
$

As is done for the Green's function, we approximate $\ket{\Psi_0}$ by the optimized dCC ground state $\ket{\Psi^{\mathbf{K}}_{\text{dCC}}}$ and $\ket{\Psi_m}$ by the tangent-space excited states $\ket{\Psi^{\mathbf{K}}_{m}}$; as noted above, only excited states within the ground-state momentum sector contribute, and in practical calculations we replace the delta functions by Lorentzians of finite width $\eta$ for presentation.

\subsection{Finite-temperature response from the tangent space}
We now extend the tangent-space response framework to finite temperature, where the same momentum-resolved description carries over.

\subsubsection{Electron-addition retarded Green's function}
For electron addition into a thermally populated phonon bath, the relevant quantity is the finite-temperature electron-addition Green's function,
\begin{equation}
    G_T(\mathbf{k},t) = -\mathrm{i}\theta(t) \text{Tr}_{\mathrm{ph}} [\hat{\rho}_\mathrm{ph} a_{\mathbf{k}}(t) a^\dagger_\mathbf{k}(0)],
\end{equation}
where $\hat{\rho}_\mathrm{ph} = e^{-\beta\hat{H}_\mathrm{ph}}/{Z_\mathrm{ph}}$ is the thermal phonon density operator and $Z_\mathrm{ph} = \text{Tr}_{\mathrm{ph}}[e^{-\beta \hat{H}_\mathrm{ph}}]$ is the corresponding partition function. Physically, this correlator describes the formation and propagation of a polaron following injection of an electron of momentum $\mathbf{k}$ into a thermal phonon ensemble. 
A direct evaluation of the phonon trace in a number-state basis would require an explicit occupation cutoff and a rapidly growing enumeration of thermal configurations. The present framework avoids this difficulty by combining the coherent-state structure of the dCC ansatz with the fact that the correlator begins with the zero-electron vacuum. The phonon trace can then be recast as a coherent-state integral whose overlaps are evaluated analytically. 

Introducing the unnormalized coherent-state measure $d\mu(\boldsymbol \alpha) = \prod_j (2\pi \mathrm{i})^{-1}\,d\bar{\alpha_j}\, d{\alpha_j}\,e^{-|\alpha_j|^2}$, chosen to match the convention used for the response manifold, and inserting the corresponding resolution of identity, $\mathbb{I} = \int d\mu(\boldsymbol \alpha) \ket{ \boldsymbol \alpha}\bra{\boldsymbol \alpha}$, the Green's function may be written as,
\begin{equation}
\begin{aligned}
    G_T(\mathbf{k},t) &= -\frac{\mathrm{i}\theta(t)}{Z_{\text{ph}}} \int d\mu(\boldsymbol \alpha) \bra{\boldsymbol \alpha} e^{-\beta\hat{H}_{\mathrm{ph}}}e^{\mathrm{i}\hat{H}t}a_{\mathbf{k}} e^{-\mathrm{i}\hat{H}t}a^\dagger_{\mathbf{k}} \ket{\boldsymbol \alpha}.\\
\end{aligned}
\end{equation}
Because the correlator begins from the zero-electron vacuum and the electron is created only upon injection via $a^\dagger_{\mathbf{k}}$, the thermal and backward time evolution on the bra side act entirely within the zero-electron sector, so the full Hamiltonian $\hat{H}$ reduces to the pure phonon Hamiltonian $\hat{H}_{\mathrm{ph}}$ there.

Inserting an approximate tangent-space resolution-of-the-identity over all momentum sectors and assuming our dCC states are eigenstates, we obtain the finite-temperature Green's function in the Lehmann representation as,
\begin{equation}
\begin{aligned}
\label{eq:finite_T_G}
    G_T(\mathbf{k},t)\simeq-\frac{\mathrm{i}\theta(t)}{Z_{\text{ph}}}\sum_{\mathbf{K}m}e^{-\mathrm{i}E_{m}^{\mathbf{K}} t} \int d\mu(\boldsymbol \alpha)\times \\\Big[\bra{\sigma(\beta, t)} a_{\mathbf{k}} \ket{\Psi^{\mathbf{K}}_{m}}\bra{\Psi^{\mathbf{K}}_{m}}a^\dagger_\mathbf{k} \ket{\boldsymbol \alpha}\Big],
\end{aligned}
\end{equation}
where the temperature and time dependence are absorbed into the bra as $\bra{\sigma(\beta, t)} =\bra{\boldsymbol \alpha}e^{-\beta \hat{H}_{\mathrm{ph}}}e^{\mathrm{i}\hat{H}_{\mathrm{ph}}t}$. Unlike the zero-temperature vacuum correlator, the thermal phonon average couples the injected particle to a distribution of initial phonon configurations, so that after the trace the tangent-space propagation is no longer restricted to the single momentum sector selected by the probe.  
As in the second-order ground-state energy evaluation, the overlaps in \cref{eq:finite_T_G} may be computed analytically using Gaussian integrals (Appendix~\ref{app:finite_T}).
Finally, finite-temperature spectra follow from the damped Fourier transform $G_T(\mathbf{k},\omega) = \int_0^{\infty}dt \; e^{\mathrm{i}\omega t} e^{-\eta t}G_T(\mathbf{k},t)$, corresponding to a Lorentzian broadening of width $\eta$ in frequency space.

\subsubsection{Optical conductivity}

Finite-temperature optical response can be tackled analogously to the zero-temperature case through the corresponding Kubo formalism. Using the approximate tangent-space excited states, the regular part of the optical conductivity is computed as in Ref.~\cite{prelovsek},
\begin{equation}
\label{eq:ft_oc}
\begin{aligned}
    \Re \,\sigma^{\mathrm{reg}}_{\alpha\alpha}(\omega) = \frac{\pi}{\Omega}\frac{1-e^{-\beta \omega}}{\omega Z}&\sum_{\mathbf{K}mn}e^{-\beta E^\mathbf{K}_{n}}|\bra{\Psi^{\mathbf{K}}_m}\hat{j}_{\alpha}\ket{\Psi^{\mathbf{K}}_n}|^2  \\
    &\times\delta(\omega - (E_m^\mathbf{K} - E_n^\mathbf{K})),
\end{aligned}
\end{equation}
where the partition function $Z = \text{Tr}[e^{-\beta \hat{H}}]$ is evaluated approximately within the tangent space, $Z \approx \sum_{\mathbf{K}n} e^{-\beta E^{\mathbf{K}}_{n}}$. 
The real part of the optical conductivity additionally includes a zero-frequency component $2\pi D_{\alpha\alpha}\delta(\omega)$, where $D_{\alpha\alpha}$ is the Drude weight, also referred to as the charge stiffness or ballistic component \cite{Prelovsek1997, Kohn1964}. 
This is computed via the formula,
\begin{equation}
\label{eq:drude_ft}
\begin{aligned}
    D_{\alpha\alpha}(\beta)& = \frac{1}{2}\langle \hat{\tau}_{\alpha\alpha}\rangle_\beta\\
    &- \sum_{\mathbf{K}}\sum_{n}\sum_{m \neq n}\frac{e^{-\beta E_n^{\mathbf{K}}}}{Z}\frac{|\langle \Psi^\mathbf{K}_m|\, \hat{j}_\alpha |\Psi^\mathbf{K}_n\rangle|^2}{E^{\mathbf{K}}_m -E^{\mathbf{K}}_n},
\end{aligned}
\end{equation}
where the second term, whose sums over the eigenstates $n$ and $m$ run over all momentum sectors, is commonly referred to as the paramagnetic term. The stress or diamagnetic operator is $\hat{\tau}_{\alpha\alpha} = \partial^2\hat{H}/\partial A_\alpha ^2|_{A_\alpha=0}$, with $A_\alpha$ an external electromagnetic vector potential introduced to the carrier hopping through the Peierls substitution \cite{prelovsek}.

Although the framework is in principle applicable to arbitrary temperature and remains systematically improvable through balanced truncations of the tangent-space hierarchy, its practical accuracy at low truncation order is expected to be best in the low- to moderate-temperature regime, where the response is dominated by the low-lying states that the manifold represents faithfully; the high-energy states that acquire appreciable thermal weight at higher temperatures lie outside this class.

\begin{figure*}[t]
\includegraphics[width=\linewidth]{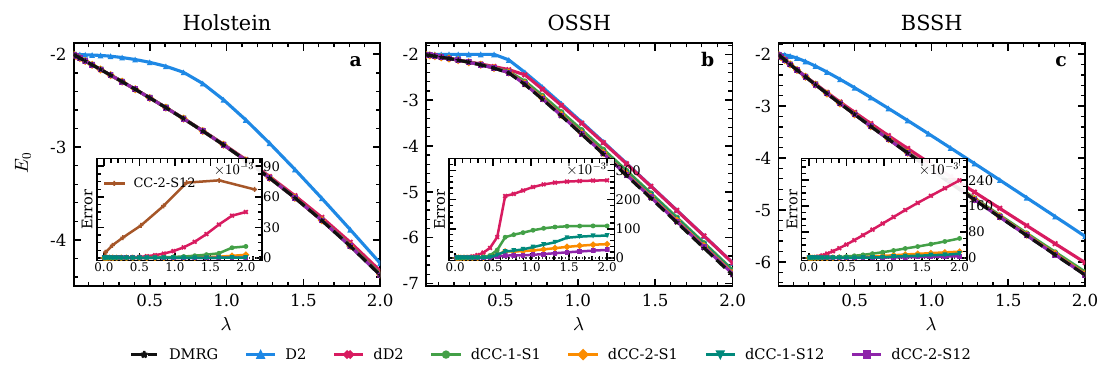}
\caption{\label{fig:1d_ground} \textbf{Ground-state energy as a function of dimensionless electron--phonon coupling strength ($\lambda$) for one-dimensional 32-site models.} (a) Holstein ($t = \omega = 1$), (b) OSSH ($t=1, \omega = 0.5$), and (c) BSSH ($t = \omega = 1$) models. The insets show the absolute error attained at each coupling strength. The CC-2-S12 result was taken from Ref.~\cite{Yang2024}, which used conventional projective CC equations.
}
\end{figure*}

\subsection{Scaling and implementation}
The computational cost of the tangent-space response is set by the dimension $D$ of the projected response manifold, summarized in \cref{tab:response_scaling}. At the $(1,1)$ level, the coupled electron--one-phonon sector dominates and $D = \mathcal{O}(N^2)$; enlarging the phonon sector to second order, as in $(2,1)$, adds only $\mathcal{O}(N^2)$ two-phonon vectors, while the coupled electron--two-phonon sector raises the dimension to $\mathcal{O}(N^3)$ in the $(1,2)$ and $(2,2)$ manifolds. The dynamical quantities considered in this work follow either from resolvents of the projected Hamiltonian or from thermal traces over the manifold, and thus admit iterative evaluation whose cost is dominated by dense matrix--vector products of $\mathcal{O}(D^2)$, for instance through continued-fraction Lanczos at zero temperature and finite-temperature Lanczos methods \cite{prelovsek} at $T>0$. At the system sizes considered in this work, we instead solve the generalized eigenvalue problem in \cref{eq:generalev} by full diagonalization at $\mathcal{O}(D^3)$ cost.

\begin{table}[!h]
\caption{\label{tab:response_scaling}
Computational scaling of the dCC-1-S1-$(n,m)$ tangent-space response for model systems with $N$ lattice sites, where $n$ and $m$ denote the highest retained orders of the phonon and coupled electron--phonon response sectors. $D$ is the dimension of the response manifold, and ``Mat-vec'' the cost of one dense matrix--vector product in iterative evaluations; the full diagonalization used in this work scales as $\mathcal{O}(D^3)$.
}
\begin{ruledtabular}
\renewcommand{\arraystretch}{1.15}
\begin{tabular}{cccc}
Response space & Additional response sector & $D$ & Mat-vec \\
\colrule
$(1,1)$ & None & $\mathcal O(N^2)$ & $\mathcal O(N^4)$ \\
$(2,1)$ & Two-phonon & $\mathcal O(N^2)$ & $\mathcal O(N^4)$ \\
$(1,2)$ & Electron--two-phonon & $\mathcal O(N^3)$ & $\mathcal O(N^6)$ \\
$(2,2)$ & Full second-order manifold & $\mathcal O(N^3)$ & $\mathcal O(N^6)$ \\
\end{tabular}
\end{ruledtabular}
\end{table}

\section{Results and Discussion}\label{sec:results}
We now benchmark the dCC hierarchy on ground states and dynamics, from lattice models to the \textit{ab initio} setting; computational details are provided in Appendix~\ref{app:comp_details}.

\subsection{Ground states}\label{sec:ground}
We begin with the ground-state properties of prototypical polaron models spanning local, bond, and long-range couplings, progressing from one- and two-dimensional lattices to the continuum Fr\"ohlich model.

\begin{figure*}[t]
    \includegraphics[width =\linewidth]{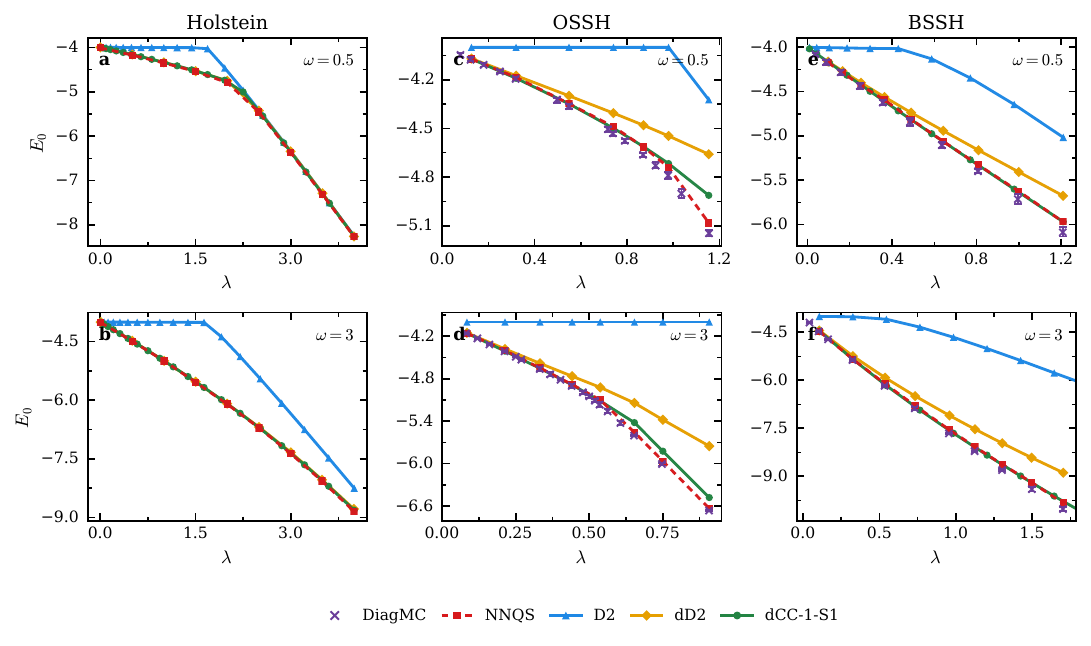}
    \caption{\label{fig:2d_ground} \textbf{2D ground states in the adiabatic $\boldsymbol\omega \mathbf{= 0.5} $ (anti-adiabatic $\boldsymbol \omega \mathbf{= 3.0}$) regime for the (a,b) Holstein (c,d) OSSH, and (e,f) BSSH models.} The D2 and dCC-1-S1 results are obtained on a $20\times20$ lattice, while the NNQS and DiagMC data digitized from Refs.~\cite{mahajan24, zhangssh} are obtained on a $10\times10$ lattice and directly in the thermodynamic limit, respectively. In accordance with previous work, we use the dimensionless coupling constant $\lambda = \frac{g^2}{2\omega t}$ for the Holstein and BSSH models and $\frac{g^2}{\omega t}$ for the OSSH model.
    }
\end{figure*}
\subsubsection{One-dimensional lattice models}

\Cref{fig:1d_ground} shows the ground-state energy obtained at various levels of dCC theory for the 32-site Holstein, BSSH, and OSSH models across a range of electron--phonon coupling strengths compared with DMRG calculations. We use the standard dimensionless coupling parameter $\lambda = {g^2}/({2\omega t})$ for the Holstein and BSSH models, and $\lambda = {g^2}/{\omega t}$ for the OSSH model. These definitions compare the phonon-mediated interaction scale with the electron's bare kinetic energy; the factor-of-two difference reflects that the OSSH hopping couples to two neighboring on-site oscillators.

In all models, we sweep the ground-state energy up to coupling strength $\lambda = 2$, which is sufficient to cover the weak-, intermediate-, and strong-coupling regimes.
Ground-state energies obtained for the dCC wavefunction are formally variational upper bounds to the lowest-energy eigenstate within each crystal momentum sector $K$; we present here the lowest energy over all momentum sectors, i.e., the ground-state energy. Across all levels of dCC theory, the Holstein and BSSH ground states remain in the $K=0$ sector throughout the considered coupling range, while the OSSH ground state undergoes a transition from $K=0$ to $K=\pi/2$ when the electron--phonon coupling becomes comparable to the phonon frequency, in agreement with previous work \cite{MarchandPRL}.

The one-dimensional results expose the limitations of simpler variational wavefunctions. In the factorized D2 product state, the lattice cannot respond conditionally to fluctuations in the carrier position, and the mean-field ansatz approaches the correct behavior only at very strong coupling. Restoring translational symmetry via the dD2 ansatz dramatically improves the Holstein ground-state energies, but substantial errors remain for the non-local OSSH and BSSH couplings, indicating the need for explicit electron--phonon correlation beyond symmetry restoration. The importance of momentum conservation is equally apparent in the Holstein results for conventional CC-2-S12~\cite{Yang2024}: although it is the most expensive conventional CC truncation shown, it is the least accurate correlated method in this comparison. In contrast, dCC-1-S1 supplements the symmetry-restored coherent-state reference with only the leading-order electron-one-phonon excitations, yet recovers the majority of the missing correlation, and all levels of dCC theory remain close to the DMRG reference across the full coupling range in all three models.

The remaining errors at strong coupling reveal the distinct roles of the two classes of second-order excitations: the electron-two-phonon sector in dCC-1-S12 describes higher-order fluctuations of the phonon cloud conditioned on the carrier, while dCC-2-S1 introduces multimode squeezing of the phonon state, which we find especially important for the OSSH model, consistent with previous numerical studies \cite{Shi2018}. Combining the two sectors in dCC-2-S12 yields the most accurate energies, as expected from the variational hierarchy, but raises the computational scaling to $\mathcal O(N^5)$ (\cref{tab:scaling}); the $\mathcal {O}(N^3)$ dCC-1-S1 theory thus provides the most favorable compromise between accuracy and cost, and we use it for all subsequent ground-state calculations unless stated otherwise.

Beyond total energies, polaron band structures follow directly by minimizing within each momentum sector. For a Holstein model with dispersive phonons, $\omega(q) = \omega_0 + 2t_{\text{ph}}\cos(q)$, which is known to be difficult for variational wavefunctions, the first-order dCC band structure agrees closely with numerically exact Lanczos benchmarks in the thermodynamic limit and with 42-site NNQS calculations throughout the Brillouin zone, correcting the large zone-edge error of dD2 (see \cref{fig:1d_disp} in Appendix~\ref{app:addl_results}) \cite{Bonca_disp, mahajan24}.

\subsubsection{Two-dimensional lattice models}
We also apply our first-order ansatz, dCC-1-S1, to study the two-dimensional Holstein, OSSH, and BSSH models. 
While the Holstein model retains a single optical phonon mode, the OSSH and BSSH models in 2D each feature two optical phonon modes per site, allowing us to probe the performance of our ansatz in models with multiple phonon bands.  
Unlike one-dimensional systems, the 2D systems are particularly challenging for existing tensor network methods. Hence, we compare our results with state-of-the-art NNQS \cite{mahajan24} and DiagMC \cite{zhangssh} calculations in \cref{fig:2d_ground}. 

For the 2D Holstein model (\cref{fig:2d_ground} (a,b)), even the mean-field D2 wavefunction provides accurate ground-state energies in the strong-coupling limit, where the coherent form of the phonon wavefunction captures the dominant energetic effect.
Restoring translational symmetry via the dD2 ansatz already yields excellent agreement with NNQS benchmarks in both the adiabatic and anti-adiabatic regimes across all considered coupling strengths, suggesting that the translationally invariant coherent-state ansatz is sufficient to capture the dominant physics of this model.
As the dCC-1-S1 wavefunction builds upon the dD2 reference, its ground-state energies inherit this accuracy throughout.

The 2D generalizations of the OSSH (\cref{fig:2d_ground} (c,d)) and BSSH (\cref{fig:2d_ground} (e,f)) models are significantly harder than their 1D counterparts. In these settings, the simple D2 coherent-state ansatz fails to qualitatively capture the correct ground-state energy curve and remains in the non-interacting limit until strong couplings are reached. Restoring translational symmetry through dD2 recovers the correct qualitative behavior, but in contrast to the Holstein case, sizable errors remain at intermediate and strong couplings in both regimes, again indicating the importance of explicit electron--phonon correlation in models with non-local couplings.

For the 2D OSSH model, dCC-1-S1 substantially reduces the dD2 error and provides accurate results at low coupling where the ground state remains at $\mathbf{K} = (0,0)$. Although dCC-1-S1 correctly captures a ground-state transition to finite crystal momentum, the ground-state energy attains a larger error in that region compared with NNQS and DiagMC values, as expected in analogy to the 1D case where higher-order coupled excitations and squeezing effects become more significant.
For the BSSH model, in contrast, the ground-state energies predicted by dCC-1-S1 are in excellent agreement with the NNQS and DiagMC references across the entire coupling range considered.

Given the relatively large residual error of dCC-1-S1 for the 2D OSSH model at strong coupling, anticipated from the 1D results in \cref{fig:1d_ground}(b), it is worth understanding the physical significance of this regime.
The OSSH model becomes unreliable at strong coupling because large electron-induced lattice distortions violate the small-displacement approximation underlying both the linear hopping modulation and the harmonic phonon potential \cite{Nocera2021,prokof2022phonon}; the problem is compounded by a geometric constraint, since each bond distortion is the difference of the displacements of the atoms it connects, so that neighboring bonds share atoms and realizing the pattern of compressed bonds favored by the polaron can require atomic displacements exceeding the individual bond distortions.
The BSSH model instead promotes the bond distortion to an independent bosonic degree of freedom, removing this constraint by construction \cite{Carbone2021,zhang2022bond}, and remains an internally consistent description of bond-coupled polarons, although its linear hopping dependence would ultimately require nonlinear corrections at very large distortions.
The residual error of dCC-1-S1 is thus concentrated in a regime where the OSSH model itself no longer describes a physical lattice; in the better-defined strong-coupling regime of the BSSH model, dCC-1-S1 remains uniformly accurate.

The dCC wavefunction also provides direct access to real-space observables characterizing the polaron structure; definitions and results for the 2D Holstein and BSSH models are collected in Appendix~\ref{app:addl_results} and \cref{fig:observables}. The Holstein density-displacement correlation is strongly peaked at the carrier, indicating a localized lattice distortion, whereas the BSSH bond correlation is anisotropic and spatially extended, reflecting the non-local nature of the bond coupling. Consistently, the quasiparticle residue decreases and the phonon number grows more rapidly with coupling in the BSSH model, signifying stronger dressing at equal nominal coupling and demonstrating that the dCC wavefunction captures the distinct real-space structures generated by local and bond electron--phonon couplings.

\begin{figure}[!h]
    \centering
    \includegraphics[width=1\linewidth]{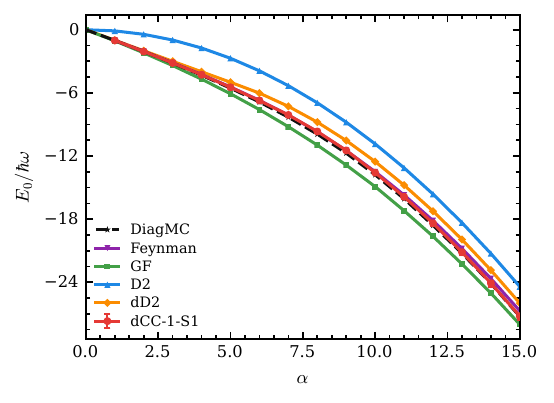}
    \caption{\textbf{Fr\"ohlich polaron ground-state energy as a function of the coupling strength $\alpha$.} Estimated continuum dCC ground-state energies are compared against DiagMC benchmark energies \cite{Hahn2018}, Feynman's path-integral solution \cite{Lu1992}, a continuum adaptation of the dD2 ansatz \cite{baumgartendd2}, the Green's function (GF) approach \cite{BartolomePRL2022}, and the D2 (Landau--Pekar) ansatz evaluated on a grid. 
    }
    \label{fig:frohlich_ground}
\end{figure}

\begin{figure*}[t]
\includegraphics[width=0.8\linewidth]{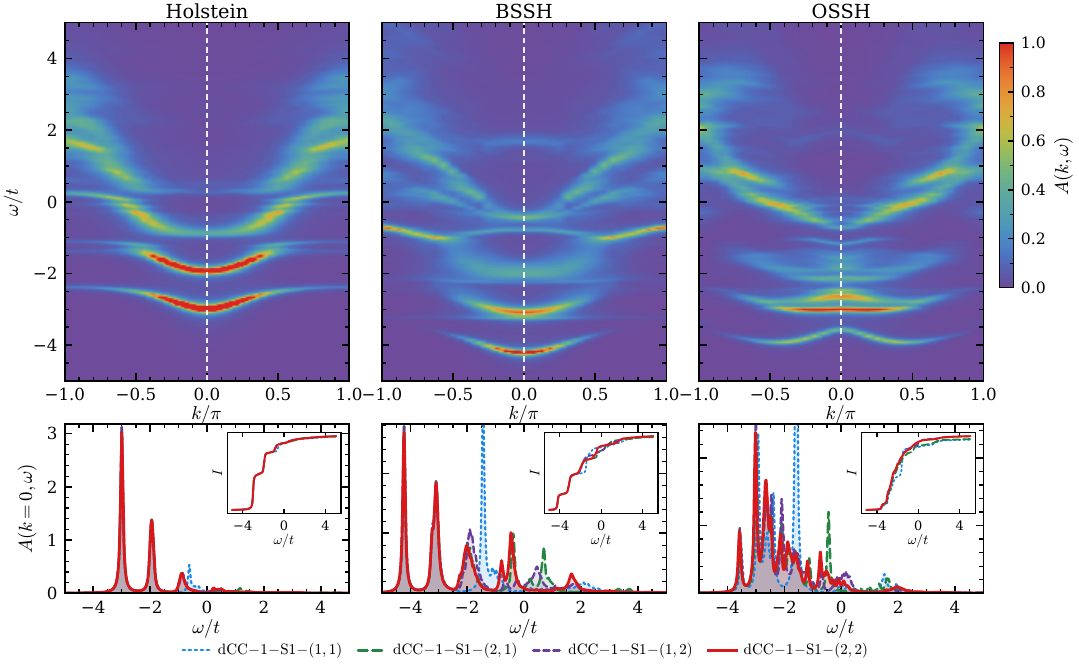}
\caption{\label{fig:1d_spectral_zt}\textbf{Zero-temperature spectral functions computed at the dCC-1-S1-$(n,m)$ level of theory for 32-site 1D lattices, where $\lambda = 1$ for all models.} Lorentzian broadening of $\eta = 0.05$ is used, and the raw spectra are interpolated for presentation. The cumulative integrated spectral weight ($I$) of the $(2,2), (2,1), (1,2), (1,1)$ levels of excited-state theory is shown in the inset.
}
\end{figure*}

\subsubsection{Three-dimensional Fr\"ohlich model}\label{sec:frohlich}
We also apply dCC-1-S1 to the three-dimensional continuum Fr\"ohlich model across the coupling range $1\leq \alpha \leq 15$. 
Unlike the lattice models considered above, the Fr\"ohlich model possesses a long-range coupling that is singular at $\mathbf{q}=0$ and lacks a finite first Brillouin zone. 
To apply our dCC approach, we consider a {\it discretized} Fr\"ohlich model defined by a periodic volume $\Omega=L^3$ and momentum cutoff $k_\text{max}$, recovering the continuum through the ordered double limit $\Omega\rightarrow \infty$ followed by $k_\text{max} \rightarrow \infty$; the formally singular $\mathbf{q}=0$ coupling and residual discretization errors are treated as described in Appendix~\ref{app:Frohlich}.
While the raw energy on each grid remains a strict variational upper bound for the discretized model, the extrapolated continuum energy is not a strict variational estimate. However, as our finite-size and discretization corrections are well-behaved, we expect the final dCC energy to closely approximate the true variational dCC energy in the continuum limit.

In \cref{fig:frohlich_ground}, we compare the first-order dCC ground-state energies with several established results for the Fr\"ohlich model. At 12 of the 15 couplings considered, the dCC estimates are closer to the DiagMC benchmark energies than Feynman's variational path-integral solution \cite{Feynman1955, Lu1992} is. 
Feynman's result is marginally more accurate for the couplings $\alpha = 7$--$9$. 
We observe that the dCC ground-state energy remains of excellent quality across the coupling range, and the largest error occurs in the crossover region, where neither a perturbative nor an adiabatic description is adequate. 
Outside the crossover regime, the one-phonon correlated excitations (\cref{eq:T1ep}) recover most of the residual correlation absent in dD2, yielding more accurate results in both weak and strong coupling regimes. 

\subsection{Zero- and finite-temperature spectral functions}\label{sec:res_dyn}
We now investigate how well dynamical features are represented by fluctuations around the optimized dCC-1-S1 ground state, using the tangent-space response formalism of \cref{sec:dynamics}, at both zero and finite temperature.

\begin{figure*}[t]
\includegraphics[width=0.75\linewidth]{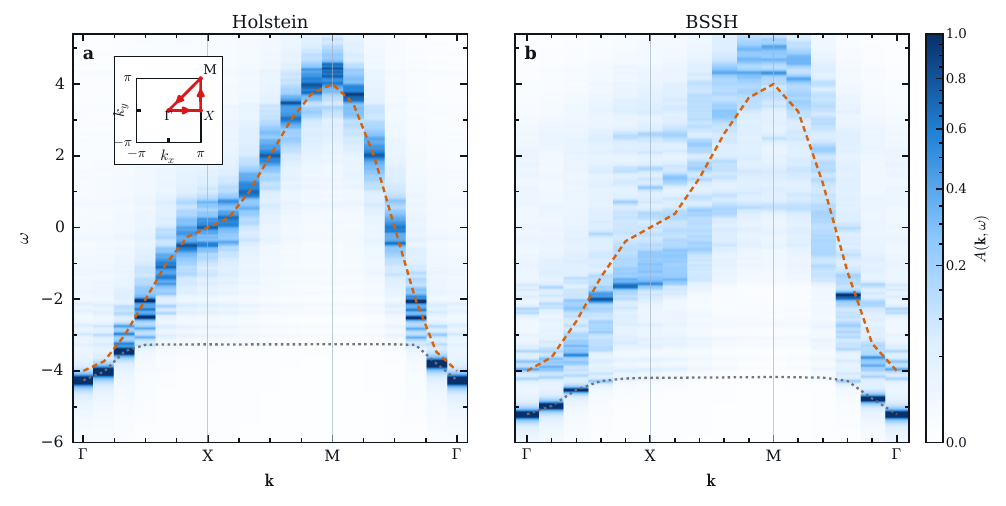}
\caption{\label{fig:2d_spec-zt}\textbf{Zero-temperature spectral function obtained from dCC-1-S1-(1,1) for a $\mathbf{12\times12}$ 2D Holstein and $\mathbf{10\times10}$ BSSH lattice ($g=1$).} Spectral function plotted along the high-symmetry path with tight-binding dispersion (dashed orange) and ground-state polaron band (dotted grey) lines overlaid for the a) Holstein and b) BSSH models. The colormap represents the normalized intensity shared between both models.
}
\end{figure*}

\subsubsection{Spectral functions at $T=0$ in one dimension}
We first consider the electron-addition spectral function, which probes the formation of a polaron upon injecting an electron with definite crystal momentum into the phonon vacuum. Unlike ground-state energies, this quantity tests not only the quality of the optimized variational ground state but also the ability of the tangent-space manifold to accurately represent the excited electron--phonon states reached after the injection process.

\cref{fig:1d_spectral_zt} shows zero-temperature spectral functions for the Holstein, OSSH, and BSSH models on 32-site one-dimensional lattices at intermediate coupling ($\lambda = 1$). Across all three coupling types, the dCC tangent-space spectra are consistent with the dominant momentum-resolved features reported in NNQS calculations~\cite{mahajan24}, including the lowest polaron branch and the higher-energy satellite features. Here, we have the advantage that we need not converge on the number of hidden neurons, a procedure that complicates physical interpretation; instead, we work with a fixed set of variational parameters, each of which is tied to a physical characteristic of the underlying dCC wavefunction.

The comparison between response manifolds illustrates how the hierarchy systematically improves the spectra. At first order, dCC-1-S1-(1,1), the response space contains electronic, phononic, and coupled electron-one-phonon fluctuations (see Appendix~\ref{app:tangent}), which suffice for the ground-state quasiparticle branch and one-phonon sidebands but not for multi-phonon satellites: too few final states are supported after electron injection, and spectral weight artificially concentrates into a few excitation channels, producing the spikes seen in \cref{fig:1d_spectral_zt}. This is most pronounced at strong coupling, where multi-phonon processes become important, so the apparent differences among the models at the same nominal $\lambda$ reflect their actual degree of phonon dressing rather than an intrinsic dependence of the response hierarchy on the model. Enlarging the response space with two-phonon and coupled electron-two-phonon sectors, i.e., (2,2), opens the missing excitation pathways and redistributes the spectral weight among the correct physical states, while all orders of the response theory agree closely in the low-energy region.

\subsubsection{Spectral functions at $T=0$ in two dimensions}
Momentum-resolved dynamical calculations are substantially more challenging in two dimensions than in one dimension: DMRG and Lanczos can no longer provide high-accuracy benchmarks at affordable cost, as spectral functions require an accurate representation of a large manifold of interacting, momentum-resolved electron--phonon excited states. Although two-dimensional Holstein spectral functions have been obtained on small clusters and by cluster perturbation theory \cite{Fehske1997, Hohenadler2003, Gao2025}, systematically improvable wavefunction calculations on larger lattices for general lattice polarons remain underexplored.

\cref{fig:2d_spec-zt} compares zero-temperature electron-addition spectra of the two-dimensional Holstein and BSSH models computed in the weak-coupling regime, corresponding to the dimensionless coupling $\lambda = 0.5$ previously defined. This coupling is smaller than the $\lambda = 1$ value used to examine the response hierarchy in one dimension, so multiphonon fluctuations should be less prominent. Therefore, the dCC-1-S1-$(1,1)$ response manifold should remain qualitatively accurate; consistently, we do not observe the pronounced, artificially intense satellite peaks that appear when the first-order manifold is insufficient, as in the more strongly dressed 1D examples.

At this coupling, both models retain a recognizable electron-addition feature tracing the tight-binding free-electron dispersion, while the lowest bright features correspond to the relaxed polaron band; the principal effects of the interaction are to lower and narrow the lowest polaron band and to redistribute spectral weight among phonon-dressed excitations around the bare-electronic feature. The redistribution is substantially larger in the BSSH model. Its polaron band is more strongly bound and flattened than that of the Holstein model, and the spectral weight spans a broader range of phonon-dressed features with correspondingly reduced intensities. This behavior is consistent with the more rapid loss of quasiparticle residue and larger phonon occupation observed in \cref{fig:observables}; although computed in the anti-adiabatic regime, that comparison shows the same qualitative contrast between the two coupling types.

\subsubsection{Spectral functions at $T>0$ in one dimension}

We next consider the finite-temperature electron-addition spectral function for the Holstein model, which describes adding an electron to a phonon bath in thermal equilibrium rather than to the phonon vacuum. As discussed in \cref{sec:dynamics}, all momentum sectors now contribute, and we evaluate the thermal phonon trace analytically in the coherent-state basis, as in recent work on Bose polarons \cite{DzostjanCoherentFT}.

\begin{figure}[tb]
\includegraphics[width =\linewidth]{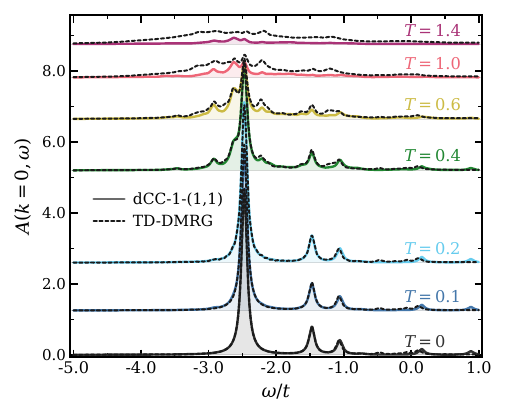}
\caption{\label{fig:1d_ft} \textbf{Finite-temperature spectral function of six-site one-dimensional Holstein model.} We used ${g = t=} \omega =1$, probe-electron momentum $\mathbf{k = 0}$ and a Lorentzian broadening of $\eta = 0.05$ for presentation. 
}
\end{figure}

\Cref{fig:1d_ft} shows the finite-temperature spectral function for a six-site Holstein model (with $g = 1$), benchmarked against finite-temperature TD-DMRG calculations. We chose this six-site system to enable comparison with previous finite-temperature Lanczos results of Bon\v{c}a \textit{et al.} \cite{BoncaFTHolstein}. While the peak positions and thermally activated features are in excellent qualitative agreement, their low-temperature spectrum exhibits a substantially stronger suppression of the quasiparticle peak at $T =0.2$ than both our dCC and TD-DMRG calculations. This suppression is difficult to reconcile with the thermal occupations: at this temperature the zero-phonon sector retains around $96\%$ of the thermal weight, suggesting only a weak suppression of the quasiparticle peak. We therefore use our TD-DMRG for the quantitative comparison in \cref{fig:1d_ft}.

We observe remarkable agreement with the TD-DMRG reference at low to intermediate temperatures. At low temperature, the response is dominated by the sharp polaronic features present in the zero-temperature spectrum. As the temperature is increased, the ground-state quasiparticle peak is suppressed as the thermal weight of the phonon vacuum, $1/Z_{\mathrm{ph}}$, decreases, and spectral weight is redistributed over a broader frequency window. At high temperature, the incompleteness of the underlying linear-response manifold is accentuated, resulting in a depletion of spectral weight relative to the TD-DMRG reference, even though the spectrum is already comparatively broad and weakly structured in this regime.

A notable qualitative difference from the zero-temperature spectra is the appearance of spectral weight below the main quasiparticle peak: because the initial phonon ensemble contains occupied modes, electron addition can be accompanied by the removal or rearrangement of thermal phonons, producing transitions at lower energies, consistent with previous finite-temperature Lanczos, cumulant, and mixed quantum-classical studies \cite{BoncaFTHolstein, NguyenMQC, Robinson2022}.
\Cref{fig:1d_ft} thus serves both as a benchmark confirming that the analytic coherent-state trace captures the leading temperature-dependent redistribution of spectral weight, and as an illustration of the low- to intermediate-temperature regime where the present implementation is expected to be accurate.

\subsubsection{Spectral functions at $T>0$ in two dimensions}
Unlike the one-dimensional benchmark, two-dimensional finite-temperature spectra are not practically accessible with conventional TD-DMRG algorithms at the system sizes considered. 
Mapping of the two-dimensional lattice to a tractable matrix-product state introduces long-range couplings not present in the short-range polaron models considered, and yields much larger spatial entanglement. 
These factors, combined with the thermal purification and real-time propagation necessary to obtain finite-temperature spectra, rapidly increase the required bond dimension and make calculations of 2D spectra prohibitively expensive. 
The unbounded local phonon Hilbert space only compounds these difficulties. 
Consequently, finite-temperature DMRG calculations of polaron spectral functions have largely been limited to one-dimensional chains \cite{jansen2020FT}. 
In contrast, by evaluating the thermal phonon trace analytically and working directly with momentum-resolved response states, our dCC approach enables momentum-resolved finite-temperature spectra on a $10\times10$ two-dimensional lattice.

\begin{figure}[tb]
    \centering
    \includegraphics[width=\linewidth]{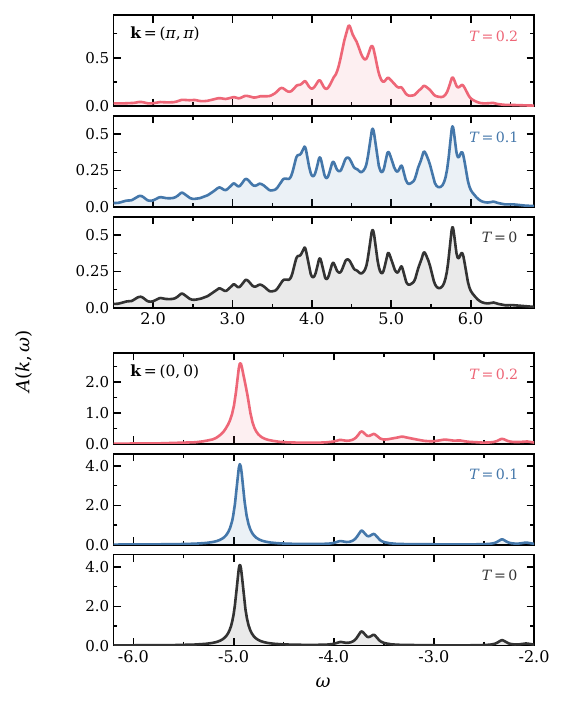}
    \caption{\textbf{Finite-temperature electron-addition spectral functions at $\mathbf{k} = (0,0)$ and $\mathbf{k} = (\pi,\pi)$ for the two-dimensional Holstein model on a $10\times 10$ lattice}. These were obtained at the dCC-1-S1-(1,1) level of theory. Slices at $\mathbf{k} = (0,0)$ (bottom) and $\mathbf{k} = (\pi,\pi)$ (top) are shown for $g = 1.8$ ($t=\omega=1$) and $T=0, 0.1, 0.2$ in units of the phonon frequency, with a Lorentzian broadening of $\eta = 0.05$.}
    \label{fig:2d_ft}
\end{figure}

\cref{fig:2d_ft} shows the resulting Holstein spectra at representative momenta $\mathbf{k} = (0,0)$ and $\mathbf{k} = (\pi,\pi)$. Here, we use $g=1.8$ to obtain appreciable temperature dependent spectral redistribution while remaining in the weakly dressed regime expected to be described well by the first-order response approach. At $\mathbf{k} = (0,0)$, the zero-temperature spectrum is dominated by a sharp peak corresponding to the polaron ground state, which is subsequently broadened and suppressed as temperature increases. Near the Brillouin zone corner at $\mathbf{k} = (\pi,\pi)$, we observe more significant temperature effects with the suppression of discrete peak structure and the concentration of spectral weight into a central feature as the system warms.

\begin{figure}[tb]
\includegraphics[width =0.9\linewidth]{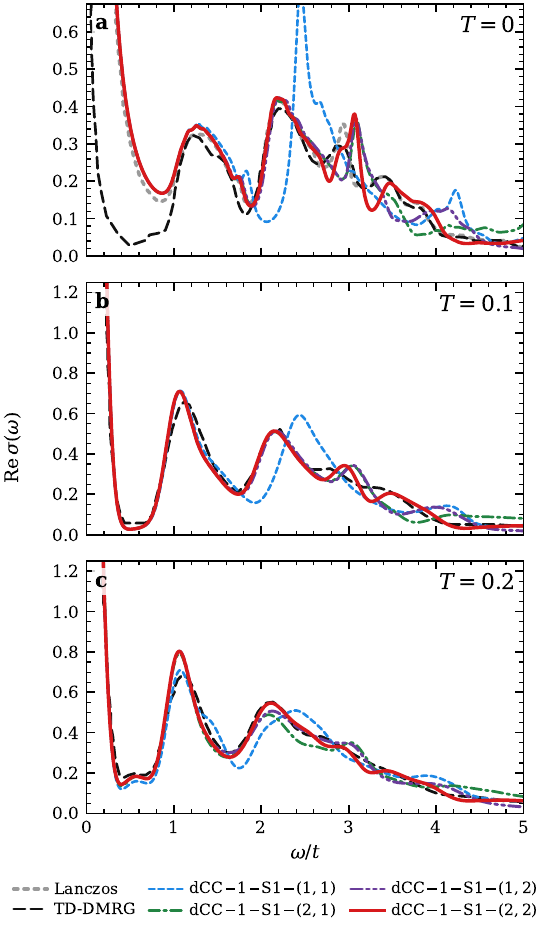}
\caption{\label{fig:1d_oc} \textbf{Finite-temperature optical conductivity of the one-dimensional 32-site Holstein model ($\lambda =1$).} We compared our results with recent TD-DMRG ($N=40$ site) calculations \cite{jansen2022}. A Lorentzian broadening of $\eta = 0.08$ is used for the zero-temperature results and a Gaussian broadening of $\eta = 0.1/(4\pi)$ for consistency with the benchmark results. We note that Lanczos benchmarks are available only at zero temperature, and we thus include them only in panel (a).
}
\end{figure}

\subsection{Zero- and finite-temperature optical conductivity}\label{subsec:optcond}
We next consider the optical conductivity, which probes a different aspect of polaron dynamics than the electron-addition spectral function and provides a more stringent test of the excited-state basis.

\subsubsection{One dimension}
\Cref{fig:1d_oc} shows the real part of the optical conductivity for the 1D Holstein model with 32 sites at intermediate coupling ($\lambda = 1$) at several temperatures, compared with Lanczos (only at $T=0$) and TD-DMRG calculations \cite{jansen2022}. Here we work directly in the frequency domain, which allows us to handle the regular component and zero-frequency singular part separately. This avoids the difficulties encountered with real-time methods (e.g., TD-DMRG), which struggle to accurately represent the low-frequency region due to the requisite long simulation times and time-step resolution. 

From \cref{fig:1d_oc}(a), all response manifolds shown agree significantly better with the zero-temperature Lanczos results in the low-frequency regime than the TD-DMRG conductivities do, and they closely reproduce the Lanczos line shape of the regular contribution. Importantly, the hierarchy systematically improves the predicted line shape at higher frequencies, with higher orders correcting the artificial accumulation of optical spectral weight, as previously observed for the spectral function.

\begin{figure*}[!ht]
\includegraphics[width =0.75\linewidth]{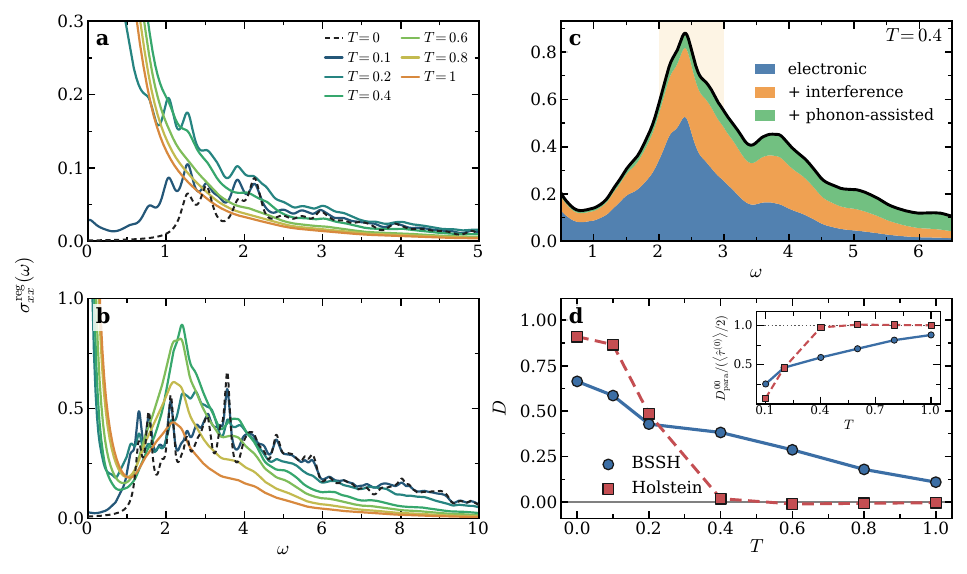}
\caption{\label{fig:2d_oc} \textbf{Optical conductivity of the two-dimensional Holstein and BSSH models using dCC-1-S1-(1,1).} a) $12\times12$ Holstein model and b) $10\times 10$ BSSH model real part of the optical conductivity $\sigma_{xx}$ ($g=t = \omega = 1$). The plots show the regular part of the optical conductivity, $\sigma_{xx}^{\text{reg}}$, as a function of temperature. c) shows the regular part of the optical conductivity at $T=0.4$ decomposed into electronic, interference, and phonon-assisted current contributions. A Lorentzian broadening of $\eta = 0.1$ is used, and the shaded band highlights the absorption peak discussed in the text. d) shows the Drude weight of the two models, with the inset showing the electronic paramagnetic fraction to the electronic stress term.
}
\end{figure*}

From \cref{fig:1d_oc}(a--c), we see that the real part of the conductivity is dominated by features clustered around $\omega/t = 1,2 $ in agreement with previous studies that attribute the former to one-phonon emission processes \cite{jansen2022, fratini}. The most pronounced temperature effect is an increase in low-frequency features, driven by the thermal population of eigenstates with small energy differences. This produces a finite regular contribution at low frequencies, which offsets a decay in the zero-frequency singular contribution $D_{\alpha\alpha}$, as evidenced by a persistent zero-frequency value of the optical conductivity with increasing temperature. Consistent with previous work \cite{jansen2022, fratini}, we find an enhancement of the existing optical conductivity peaks due to contributions from states populating non-zero crystal momentum sectors at finite temperatures.

This comparison also highlights an important subtlety within the response hierarchy. At low temperatures, the balanced and partially enlarged linear response manifolds yield similar results. However, as temperature increases, the unbalanced dCC-1-S1-$(1,2)$ and dCC-1-S1-$(2,1)$ manifolds can separate strongly because of their asymmetric treatment of phonon and coupled electron--phonon channels, and do not themselves constitute a finite-temperature convergence sequence. We additionally emphasize that, although the first-order response space does not yet quantitatively predict converged finite-frequency line shapes, it reproduces the qualitative structure obtained from the higher-order response manifolds.

\subsubsection{Two dimensions}
Finite-temperature optical conductivity remains particularly challenging in two dimensions.
Recent mixed quantum--classical Green--Kubo calculations have been applied to large two-dimensional {\it ab initio} systems \cite{jiang2026first}, but a systematically improvable fully quantum treatment at comparable scales remains unavailable and mixed quantum--classical approaches still struggle at low temperatures. 
While the electron-addition spectral function involves the propagation of a single-electron injection, the optical conductivity involves current-operator matrix elements between pairs of eigenstates in every thermally populated crystal momentum sector. 
An adequate description therefore requires either an accurate treatment of many thermally accessible excited states, whose number grows rapidly with system size and temperature, or the extraction of the conductivity from long-time current correlation functions. 
Both routes become increasingly difficult in two dimensions; consequently, simultaneously handling finite-temperature, large two-dimensional lattices and real-frequency optical response remains largely inaccessible to existing methodologies \cite{jansen2022, Goodvin2011, Mischenko2003}.

In \cref{fig:2d_oc}(a,b), we show the dCC-1-S1-(1,1) framework applied to computation of the optical conductivity of the 2D Holstein and BSSH models at the dimensionless coupling $\lambda = 1/2$. At low temperatures, the regular part of the Holstein response is concentrated near the lower end of the one-phonon continuum and decays relatively rapidly with increasing frequency. The BSSH response instead extends over a much broader frequency range, forming a broad maximum at a frequency of order $2\omega_0$ and retaining significant spectral weight at substantially higher frequencies. 

This contrast can be understood from the leading one-phonon absorption processes derived in Appendix~\ref{app:oc_pert}. For the nearest-neighbor dispersion $\epsilon_{\mathbf q}$, the one-phonon excitation energies $\Delta_\mathbf q = \omega_0 + (\epsilon_{\mathbf q} - \epsilon_0)$, an electron recoiling from the band minimum to momentum $\mathbf q$ while emitting a phonon of momentum $-\mathbf q$, span the same window, $\omega_0 \leq \Delta_\mathbf q \leq \omega_0 + 8t$, in both models; the qualitatively different distributions of optical weight within this window instead reflect the different structures of the electron--phonon coupling, whose leading-order contributions scale as $\sin^2 (q_x)\,\Delta_\mathbf q^{-3}$ for the Holstein model and as $\sin^2 (q_x/2)\,\Delta_\mathbf q^{-1}$ for the BSSH model.

In the Holstein model, the local interaction does not enter the current operator, so absorption into the one-phonon continuum proceeds only indirectly through interaction-induced dressing of the ground state; the resulting $\Delta_\mathbf q^{-3}$ dependence, together with a $\sin(q_x)$ matrix element that vanishes at both $q_x = 0$ and the Brillouin-zone boundary, strongly suppresses the response at higher frequencies. In the BSSH model, the bond coupling instead contributes a phonon-assisted term directly to the current operator, producing contributions that decay only as $\Delta_\mathbf q^{-1}$ with a $\sin(q_x/2)$ factor that survives toward the zone edge, and hence a broader response with weaker high-frequency suppression. This perturbative analysis qualitatively accounts for the contrasting line shapes in \cref{fig:2d_oc}.

We can further gain additional insight by decomposing the observed dynamics using our nonperturbative response framework. As the BSSH interaction is linear in the bond displacement, the current operator separates exactly into an electronic and phonon-assisted piece (see~\cref{eq:current_decomp}), and the regular component decomposes accordingly as $\sigma_\text{reg}(\omega) = \sigma^{00}_\text{reg}(\omega) + \sigma^{01}_\text{reg}(\omega) + \sigma^{11}_\text{reg}(\omega)$, where the terms correspond to contributions from the pure electronic current, interference term, and phonon-assisted piece of the current. \cref{fig:2d_oc}(c) shows this decomposition at $T=0.4$, revealing that the direct phonon-mediated and interference terms arising from the bond coupling play a substantial role in giving rise to the observed line shape.

The development of the finite-temperature features reflects which initial states contribute to the thermal response: the conductivity is a Boltzmann-weighted sum over all total-momentum sectors (see \cref{eq:ft_oc}), and because strong hybridization with the one-phonon continuum caps the polaron bandwidth at approximately $\omega_0$ above its minimum, a large number of momentum states cluster near the band edge and collectively outweigh the zone-center contributions despite their small individual Boltzmann factors. The finite-temperature maximum for $T\geq 0.4$ accordingly settles near the absorption maximum of these band-edge sectors, in the $\omega \simeq 2$--$3$ region (see \cref{fig:2d_oc_sector}, Appendix~\ref{app:oc_pert}).

The same decomposition of the current contributions clarifies the contrasting behavior of the Drude weight $D_{xx}$. Decomposing $D_{xx}$ for the BSSH model as 
\begin{equation}
  D_{xx}=\frac{1}{2}\big\langle\hat{\tau}^{(0)}_{xx} +\hat{\tau}^{(1)}_{xx}\big\rangle-\big( D^{00}_{\text{para}} +D^{01}_{\text{para}} + D^{11}_{\text{para}}
   \big),
\end{equation}
we find that the phonon-assisted contribution is nearly self-canceling: across the considered temperatures, $\frac{1}{2}\langle \hat{\tau}^{(1)}_{xx}\rangle - D^{01}_\text{para} -  D^{11}_\text{para}$ remains a small fraction of the total diamagnetic weight, so the phonon-assisted current generates substantial finite-frequency spectral weight while consuming almost exactly its own contribution to the optical sum rule. This leaves $D_{xx}$ governed by the competition between the electronic stress $\langle \hat{\tau}^{(0)}_{xx}\rangle$ and the electronic relaxation term $D^{00}_\text{para}$; in the Holstein model, the Drude weight is governed entirely by this competition.
The two models differ sharply in how the electronic paramagnetic term exhausts its sum-rule reservoir (see \cref{fig:2d_oc}(d)): the Holstein ratio $D_\text{para}^{00}/ (\langle \hat{\tau}^{(0)}_{xx}\rangle/2)$ grows rapidly with temperature, effectively extinguishing the Drude weight by $T=0.4$, whereas the corresponding BSSH ratio begins at a higher magnitude but saturates more slowly.
Consistently, the contributions to the Drude weight from each momentum sector ($D_\mathbf{K}$, \cref{eq:drude_ft} restricted to one $\mathbf{K}$ sector) show no analog of the Holstein collapse: the ratio $D_\mathbf{K}/(\langle\hat{\tau}\rangle_{\mathbf{K}}/2)$ for the band-edge states that dominate the thermal ensemble remains comparable to its $\Gamma$-point value, so thermal population replaces the ground state with states of similar ballistic character, reordering the Holstein and BSSH Drude weights at moderate temperature despite the stronger BSSH dressing at the same coupling.

\subsection{Towards \textit{ab initio} simulations}\label{sec:ab_initio}

\begin{figure}[tb]
\includegraphics[width = 1\linewidth]{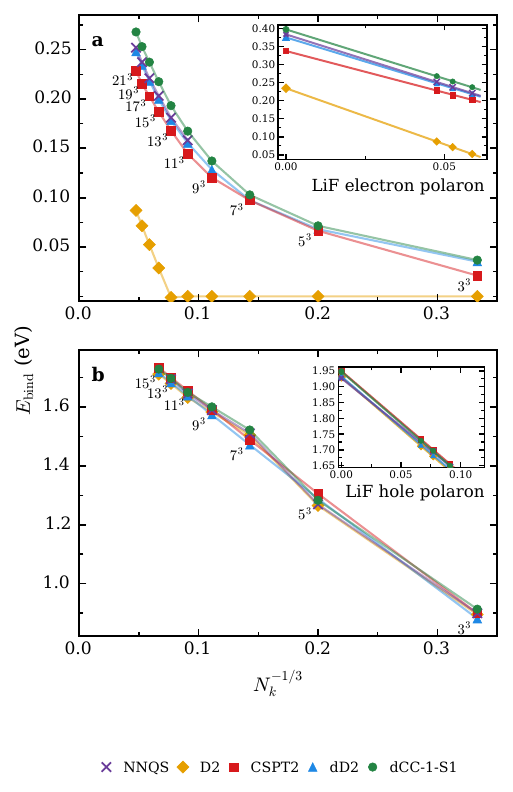}
\caption{\label{fig:LiF} \textbf{Predicted binding energies for the LiF electron and hole polarons.} The $\mathbf{k}$-mesh sizes presented range from $3^3$ to $21^3$ for the electron-polaron and $3^3$ to $15^3$ for the hole-polaron. A 3-point linear extrapolation of the methods compared is shown in the inset, with extrapolated values reported in \cref{tab:lif_binding}. Direct comparisons should be interpreted with caution, as different calculations use different mesh sizes and interpolation procedures.
}
\end{figure}

To demonstrate that our current framework can be applied with \textit{ab initio} electron--phonon matrix elements without additional model assumptions, we evaluate the lithium fluoride (LiF) electron- and hole-polaron binding energies using the momentum-space Hamiltonian in \cref{eq:mom_ham}. LiF has become a standard benchmark for \textit{ab initio} polaron methods, having been studied extensively with the Green's function (GF) and \textit{ab initio} polaron-equation approaches of Giustino and co-workers \cite{Sio2019, Sio2019PRB, Bartolome2022, GiustinoReview}, the canonical-transformation and first-principles DiagMC approaches of Bernardi and co-workers \cite{LeeBernardi2021, Luodqmc}, and recent variational and NNQS wavefunction methods \cite{mahajan24, baumgartendd2, RobinsonAIPolaron}. 
For a sequence of uniform $\mathbf{k}$-meshes ranging from size $3^3$ to $21^3$ for the electron polaron and $3^3$ to $15^3$ for the hole, we optimize the first-order dCC wavefunction and report the binding energy relative to the free-electron limit, $E_{\text{bind}} = E_{\text{free}} - E_{\text{pol}}$, in \cref{fig:LiF}.

Even at the system sizes considered, a linear extrapolation \cite{MakovPayne} of the binding energies obtained from the densest three meshes yields excellent agreement with recent state-of-the-art values, as summarized in \cref{tab:lif_binding} \cite{baumgartendd2, mahajan24, RobinsonAIPolaron,Luodqmc}. In particular, dCC-1-S1 yields a higher binding energy and, therefore, a lower variational total energy than recent NNQS results across all considered grid sizes, and largely closes the energy gap between dD2 and DiagMC. This indicates that our present ansatz may be more expressive or easier to optimize than NNQS at practically accessible network depths. Our extrapolated LiF electron-polaron binding energies agree closely with published DiagMC \cite{Luodqmc} and dD2 \cite{baumgartendd2} results obtained from large grids. As the dCC extrapolation uses mesh sizes no larger than $21^3$, some residual finite-size error is expected; a comparison between dD2 extrapolations on the same meshes and dD2's thermodynamic limit (TDL) suggests that our extrapolated value remains slightly underbound relative to the true dCC TDL estimate.

\begin{table}[!h]
\caption{\label{tab:lif_binding}LiF electron- and hole-polaron binding energies in eV, reported as positive binding magnitudes. The dCC values were obtained by
extrapolating the three densest meshes in \cref{fig:LiF}. The published dD2, GF, and DiagMC calculations employ larger mesh sizes, and our estimates, obtained at smaller meshes, are expected to improve with larger system sizes.
}
\begin{ruledtabular}
\begin{tabular}{lcc}
Method&Electron&Hole\\
\hline
DiagMC \cite{Luodqmc}&0.408&2.26\footnote{We believe this number may be significantly different from other methods due to the residual bias in DiagMC. See main text for further discussion.}\\
NNQS \cite{mahajan24}\footnote{Reported NNQS values are obtained with an independent linear extrapolation using the densest three meshes from Ref. \cite{mahajan24}.} &0.383& 1.927\\
GF \cite{Bartolome2022}\footnote{Reported without additional Debye--Waller and Allen--Heine--Cardona corrections to ensure consistent comparison.} & 0.246& 1.897 \\
CSPT2 \cite{baumgartendd2} & 0.363 & 1.950\\
D2 \cite{baumgartendd2} & 0.239& 1.933 \\
dD2 \cite{baumgartendd2}&0.395&1.933\\
dCC-1-S1 (this work)&0.397&1.948\\
\end{tabular}
\end{ruledtabular}
\end{table}

For the hole polaron, our result is consistent with GF, NNQS, variational, and perturbative approaches, but remains substantially below the reported DiagMC binding energy. We note that the hole polaron lies deep in the strong-coupling regime, where the diagrammatic expansion converges most slowly and residual bias is most difficult to remove~\cite{baumgartendd2}. 
Variational ans\"atze of very different form, the coherent-state-based dD2 and dCC wavefunctions and the NNQS parameterization, agree to within roughly $0.02$~eV of one another. The published dD2 value was obtained on meshes larger than ours, making residual finite-size error an unlikely explanation for the gap.
Attributing the discrepancy entirely to missing correlation would require these very different wavefunction forms to miss nearly identical amounts of correlation energy, suggesting that part of it may instead reflect bias in the diagrammatic reference.

The first-order dCC ansatz is structurally close to coherent-state second-order perturbation theory (CSPT2) \cite{Lee2021, RobinsonAIPolaron}: both introduce one-phonon excitations above a displaced phonon vacuum, but CSPT2 determines the amplitudes perturbatively on an unprojected reference and evaluates the energy through second order, whereas dCC determines them variationally on a momentum-projected one. The variational optimization retains the coherent-state description where it is adequate (strong coupling) while capturing one-phonon corrections that CSPT2 represents only perturbatively.

\section{Conclusions and Outlook}\label{sec:conclusion}
We have introduced dCC, a translationally invariant variational coupled-cluster theory for single polarons. Three structural features make this nonperturbative framework practical: momentum projection restores the translational symmetry broken by the coherent-state reference and gives direct access to polaron energy bands; the single-carrier structure admits closed-form variational energies at cost as low as $\mathcal{O}(N^3)$; and the coherent-state reference requires no explicit phonon-number cutoff. The same excitation hierarchy that systematically improves the ground state extends, through the projected tangent-space response formalism, to spectral functions and optical conductivities at zero and finite temperature, with thermal phonon traces evaluated analytically rather than by enumerating thermally occupied phonon-number configurations. The theory thus bridges longstanding coupled-cluster ideas from electronic structure and variational ans\"atze developed for modern polaron problems.

We validated the framework across the canonical polaron models. dCC ground-state energies closely track DMRG, NNQS, and DiagMC benchmarks for the one- and two-dimensional Holstein and SSH (optical and bond) models, reproduce Lanczos polaron band structures even for dispersive phonons, and lie closer to DiagMC than Feynman's variational solution for the Fr\"ohlich model at most couplings considered. The tangent-space response reproduces Lanczos and TD-DMRG spectral functions and optical conductivities in one dimension at zero and finite temperature, and extends momentum-resolved finite-temperature spectra to two-dimensional lattices beyond the practical reach of these methods. Applied directly to \textit{ab initio} electron--phonon matrix elements, dCC yields LiF electron- and hole-polaron binding energies in agreement with state-of-the-art many-body methods, with a lower variational energy than NNQS.

Owing to its low-order polynomial scaling, the ground-state framework applies to materials-specific Hamiltonians without modification, and its integration with recently developed electron--phonon compression schemes \cite{BernardiSVD} should enable material properties to be converged to the TDL. The variational structure also invites extensions to bipolarons, where two carriers interact with a shared phonon environment, and to time-dependent variational principle algorithms for excited-state and non-equilibrium dynamics across wider parameter regimes and temperatures.

More broadly, this work demonstrates that quantum-chemical many-body methods and traditional polaron approaches have much to offer each other. The coupled-cluster formalism developed here provides a unified variational treatment of electron--phonon polaron ground states and dynamics, and we hope that related ideas may be brought to bear on other polaronic problems of modern interest, including Bose and spin polarons.

\section*{acknowledgments}
This work was supported by the startup funds from Harvard University.
The authors thank Jinghong Zhang and P.J. Robinson for insightful discussions. T.J. acknowledges support from the Gordon and Betty Moore Foundation Fellowship. H.W. would like to acknowledge Jonah Downs, Phil Che, and Andrew Yue for valuable moral support.

\section*{Data Availability}
The numerical data necessary to reproduce the included figures is publicly available at \url{https://github.com/JoonhoLee-Group/dcc_data}. 
\bibliography{apssamp}
\clearpage
\onecolumngrid
\appendix
\setcounter{figure}{0}
\renewcommand{\thefigure}{A\arabic{figure}}
\setcounter{table}{0}
\renewcommand{\thetable}{A\arabic{table}}
\counterwithout*{equation}{section}
\setcounter{equation}{0}
\renewcommand{\theequation}{A\arabic{equation}}
\section*{Appendices}
\section{Computational Details}\label{app:comp_details}
DMRG reference values are obtained with the Renormalizer package \cite{litddmrg}. Finite-temperature DMRG spectra are obtained with thermofield dynamics using a bond dimension of $700$ with a phonon cutoff of $32$ per mode for $T\geq 0.4$ and bond dimension $128$ with per-mode phonon cutoff $12$ for lower temperatures so that lineshapes are qualitatively converged. We use a time step of $\Delta t = 0.1$ with a total time of $t_\text{max}=150$ ($t_\text{max}=80$ for $T=0.2$). The dCC variational parameters were optimized by direct minimization of the energy using analytic gradients and diagonal Hessian preconditioning. Optimizations were performed in Python using either geometric direct-minimization (GDM) \cite{van_voorhis2002} implemented in a development version of Q--Chem \cite{QChem}, or a line-search limited memory BFGS (L-BFGS-LS) implementation in JAX \cite{jax2018,jaxopt}. For the \textit{ab initio } calculations, we work directly with the \textit{ab initio} band-dispersion $\epsilon_{i_\mathbf{k}}$, phonon frequencies $\omega_{\nu \mathbf{q}}$, and electron--phonon couplings $g^\nu_{ij}(\mathbf{k}, \mathbf{q})$ generated from Quantum Espresso \cite{QE2017} and EPW \cite{EPW}, retaining all six phonon branches and relevant electronic bands. 
\section{First-order Energy Expression}\label{app:first_order}
Here we provide the complete derivation of the first-order polaron energy expression at a chosen total crystal quasimomentum $\vec{K}$ using the linear-coupling electron--phonon Hamiltonian expressed in momentum space \cref{eq:mom_ham}. In the following we will denote the dD2 type contribution to the first-order wavefunction in \cref{eq:first_order_dcc} $\ket{1}$ and the single-phonon corrected piece $\ket{2}$. Beginning with the overlap, the three necessary pieces can be written as
\begin{equation}
    \begin{aligned}
        \bra{1}\ket{1} &= \sum_{dj_1j_2k_1k_2}e^{i\vec{K}d} \tilde{\zeta}^*_{k_1, j_1}\tilde{\zeta}_{k_2, j_2}\bra{0}\wick{\c1 a_{{j_1}_{{k}_1}} \c1 a^\dagger_{{j_2}_{{k}_2}}}\ket{0}_{\text{el}}e^{-ik_2d} \bra{\Theta_{q'\nu}}\ket{\Theta_{q'\nu}e^{-iq'd}}\\
        &= \sum_{dj_1k_1} e^{i\vec{K}d}\tilde{\zeta}_{k_1, j_1}^*\tilde{\zeta}_{k_1, j_1}e^{-ik_1d}S_d
    \end{aligned}
\end{equation}
\begin{equation}
    \begin{aligned}
        \bra{1}\ket{2} &= \sum_{\mathclap{\substack{dk_1k_2j_1\\j_2q_2\nu_2}}}e^{i\vec{K}d}\tilde{\zeta}^*_{k_1, j_1} \tilde{\gamma}^{j_2, \nu_2}_{k_2, q_2} \bra{0}\wick{\c1 a_{{j_1}_{{k}_1}} \c1 a^\dagger_{{j_2}_{{k}_2}}}\ket{0}_{\text{el}} e^{-ik_2d} \bra{\Theta_{q'\nu}}b^\dagger_{{\nu_2}_{\mathbf{q}_2}}e^{-iq_2d}\ket{\Theta_{q'\nu}e^{-iq'd}}\\
        &= \sum_{dk_1j_1 q_2 \nu_2}e^{i\vec{K}d} \tilde{\zeta}^*_{k_1, j_1}\tilde{\gamma}^{j_1, \nu_2}_{k_1, q_2} e^{-ik_1d}e^{-iq_2d}\Theta^*_{q_2\nu_2}S_d,
    \end{aligned}
\end{equation}
\begin{equation}
    \begin{aligned}
        \bra{2}\ket{2} &= \sum_{\mathclap{\substack{dk_1k_2j_1j_2\\q_1 q_2 \nu_1 \nu_2}}} e^{i\vec{K}d}\tilde{\gamma}^{j_1, \nu_1*}_{k_1, q_1}\tilde{\gamma}^{j_2, \nu_2}_{k_2, q_2} \bra{0} \wick{\c1 a_{{j_1}_{{k}_1}} \c1 a^\dagger_{{j_2}_{{k}_2}}}\ket{0}_{\text{el}}e^{-ik_2d}\bra{\Theta_{q'\nu}}b_{{\nu_1}_{\mathbf{q}_1}}b^\dagger_{{\nu_2}_{\mathbf{q}_2}}e^{-iq_2d}\ket{\Theta_{q'\nu}e^{-iq'd}}\\
        &= \sum_{\mathclap{\substack{dk_1j_1q_1\\q_2\nu_1\nu_2}}}e^{i\vec{K}d}\tilde{\gamma}^{j_1, \nu_1 *}_{k_1, q_1}\tilde{\gamma}^{j_1, \nu_2}_{k_1, q_2}e^{-ik_1d}e^{-iq_2d}e^{-iq_1d}\Theta^*_{q_2\nu_2}\Theta_{q_1\nu_1}S_d\\
        &+ \sum_{\mathclap{\substack{dk_1j_1 q_1 \nu_1}}}e^{i\vec{K}d}\tilde{\gamma}^{j_1, \nu_1*}_{k_1, q_1}\tilde{\gamma}^{j_1, \nu_1}_{k_1, q_1}e^{-ik_1d}e^{-iq_1d}S_d
    \end{aligned}
\end{equation}
with the total overlap computed as
\begin{equation}
    O(\Psi) = \bra{1}\ket{1} + \bra{2}\ket{2} + 2\text{Re}(\bra{1}\ket{2}).
\end{equation}
In the above, $S_d$ denotes the coherent state overlap given by
\begin{equation}
    S_d :=\exp(\sum_{q'\nu}-\frac{1}{2}(|\Theta_{q'\nu}|^2 + |\Theta_{q'\nu}|^2 - 2|\Theta_{q'\nu}|^2e^{-iq'd})) = \exp(\sum_{q'\nu}-|\Theta_{q'\nu}|^2(1 - e^{-iq'd}))
\end{equation}
for normalized coherent states. The detailed Hamiltonian matrix elements are derived below. The contributions $H_\text{el}(\Psi)$ from the electronic part of the Hamiltonian are
\begin{equation}
    \begin{aligned}
        \bra{1}\hat{H}_{\text{el}}\ket{1} &= \sum_{\mathclap{\substack{diqk_1\\
        k_2 j_1 j_2}}}e^{i\vec{K}d}\epsilon_{iq}\tilde{\zeta}^*_{k_1, j_1}\tilde{\zeta}_{k_2, j_2}\bra{0}\wick{\c1 a_{{j_1}_{{k}_1}} \c1 a^\dagger_{i_{{q}}} \c1 a_{i_{q}} \c1 a^\dagger_{ {j_2}_{{k}_2}}}\ket{0}_{\text{el}}e^{-ik_2d}\bra{\Theta_{q'\nu}}\ket{\Theta_{q'\nu}e^{-iq'd}}\\
        &= \sum_{diq} e^{i\vec{K}d}\epsilon_{iq}\tilde{\zeta}^*_{q, i}\tilde{\zeta}_{q,i}e^{-iqd}S_d
    \end{aligned}
\end{equation}
\begin{equation}
    \begin{aligned}
        \bra{1}\hat{H}_{\text{el}}\ket{2} &= \sum_{\mathclap{\substack{diqk_1 k_2\\
        j_1 j_2 q_2 \nu_2}}} e^{i\vec{K}d}\epsilon_{iq}\tilde{\zeta}^*_{k_1, j_1} \tilde{\gamma}^{j_2, \nu_2}_{k_2, q_2} \bra{0}\wick{\c1 a_{{j_1}_{{k}_1}} \c1 a^\dagger_{i_{q}} \c1 a_{i_{q}} \c1 a^\dagger_{{j_2}_{{k}_2}}} \ket{0}_{\text{el}}e^{-ik_2d} \bra{\Theta_{q'\nu}}b^\dagger_{{\nu_2}_{\mathbf{q}_2}}e^{-iq_2d}\ket{\Theta_{q'\nu}e^{-iq'd}}\\
        &= \sum_{diqq_2\nu_2}e^{i\vec{K}d}\epsilon_{iq}\tilde{\zeta}^*_{q,i}\tilde{\gamma}^{i, \nu_2}_{q, q_2}e^{-iqd}e^{-iq_2d}\Theta^*_{q_2\nu_2}S_d
    \end{aligned}
\end{equation}
\begin{equation}
    \begin{aligned}
        \bra{2}\hat{H}_{\text{el}}\ket{2} &= \sum_{\mathclap{\substack{diqk_1k_2 j_1\\j_2q_1 q_2 \nu_1 \nu_2}}} e^{i\vec{K}d}\epsilon_{iq}\tilde{\gamma}^{j_1, \nu_1*}_{k_1, q_1} \tilde{\gamma}^{j_2, \nu_2}_{k_2, q_2} \bra{0}\wick{\c1 a_{{j_1}_{{k}_1}} \c1 a^\dagger_{i_{q}} \c1 a_{i_{q}} \c1 a^\dagger_{{j_2}_{{k}_2}}}\ket{0}_{\text{el}}e^{-ik_2d}\bra{\Theta_{q'\nu}}b_{{\nu_1}_{\mathbf{q}_1}}b^\dagger_{{\nu_2}_{\mathbf{q}_2}}e^{-iq_2d}\ket{\Theta_{q'\nu}e^{-iq'd}}\\
        &= \sum_{\mathclap{\substack{diqq_1\\q_2\nu_1\nu_2}}}e^{i\vec{K}d}\epsilon_{iq}\tilde{\gamma}^{i, \nu_1*}_{q, q_1}\tilde{\gamma}^{i, \nu_2}_{q, q_2}e^{-iqd}e^{-iq_2d} e^{-iq_1d}\Theta^*_{q_2\nu_2}\Theta_{q_1 \nu_1}S_d\\
        &+ \sum_{diq q_1\nu_1}e^{i\vec{K}d}\epsilon_{iq}\tilde{\gamma}^{i, \nu_1 *}_{q, q_1} \tilde{\gamma}^{i, \nu_1}_{q, q_1} e^{-iqd}e^{-iq_1d}S_d.
    \end{aligned}
\end{equation}
The contributions from the pure-phonon part of the Hamiltonian $H_\text{ph}(\Psi)$ are computed as 
\begin{equation}
    \begin{aligned}
        \bra{1}\hat{H}_{\text{ph}}\ket{1} &= \sum_{\mathclap{\substack{dq\nu j_1 j_2\\k_1k_2}}}e^{i\vec{K}d}\omega_{q\nu} \tilde{\zeta}^*_{k_1, j_1}\tilde{\zeta}_{k_2, j_2}\bra{0}\wick{\c1 a_{{j_1}_{{k}_1}} \c1 a^\dagger_{{j_2}_{{k}_2}}}\ket{0}_{\text{el}}e^{-ik_2d}\bra{\Theta_{q'\nu}}b^\dagger_{\nu_\mathbf{q}} b_{\nu_\mathbf{q}}\ket{\Theta_{q'\nu}e^{-iq'd}}\\
        &=\sum_{dq\nu k_1 j_1}e^{i\vec{K}d}\omega_{q\nu}\tilde{\zeta}^*_{k_1, j_1}\tilde{\zeta}_{k_1, j_1}e^{-ik_1d} \Theta^*_{q\nu}\Theta_{q\nu}e^{-iqd}S_d
    \end{aligned}
\end{equation}
\begin{equation}
    \begin{aligned}
        \bra{1}\hat{H}_{\text{ph}}\ket{2} &= \sum_{\mathclap{\substack{dq\nu j_1 j_2\\k_1 k_2 q_2\nu_2}}} e^{i\vec{K}d}\omega_{q\nu}\tilde{\zeta}^*_{k_1, j_1}\tilde{\gamma}^{j_2, \nu_2}_{k_2, q_2}\bra{0}\wick{\c1 a_{{j_1}_{{k}_1}} \c1 a^\dagger_{{j_2}_{{k}_2}}}\ket{0}_{\text{el}}e^{-ik_2d}\bra{\Theta_{q'\nu}}b^\dagger_{\nu_\mathbf{q}}b_{\nu_\mathbf{q}}b^\dagger_{{\nu_2}_{\mathbf{q}_2}}e^{-iq_2d}\ket{\Theta_{q'\nu}e^{-iq'd}}\\
        &= \sum_{\mathclap{\substack{dq\nu j_1\\k_1q_2\nu_2}}}e^{i\vec{K}d}\omega_{q\nu}\tilde{\zeta}^*_{k_1, j_1}\tilde{\gamma}^{j_1, \nu_2}_{k_1, q_2}e^{-ik_1d}e^{-iq_2d}e^{-iqd}\Theta^*_{q\nu}\Theta^*_{q_2\nu_2}\Theta_{q\nu}S_d\\
        &+ \sum_{\mathclap{\substack{dq\nu j_1 k_1}}}e^{i\vec{K}d}\omega_{q\nu}\tilde{\zeta}^*_{k_1, j_1}\tilde{\gamma}^{j_1, \nu}_{k_1, q}e^{-ik_1d}e^{-iqd}\Theta^*_{q\nu}S_d
    \end{aligned}
\end{equation}
\begin{equation}
    \begin{aligned}
        \bra{2}\hat{H}_{\text{ph}}\ket{2} &= \sum_{\mathclap{\substack{dq\nu j_1 j_2\\k_1k_2q_1q_2 \nu_1\nu_2}}} e^{i\vec{K}d}\omega_{q\nu}\tilde{\gamma}^{j_1, \nu_1*}_{k_1, q_1}\tilde{\gamma}^{j_2, \nu_2}_{k_2, q_2} \bra{0}\wick{\c1 a_{{j_1}_{{k}_1}} \c1 a^\dagger_{{j_2}_{{k}_2}}}\ket{0}_{\text{el}}e^{-ik_2d}\bra{\Theta_{q'\nu}}b_{{\nu_1}_{\mathbf{q}_1}}b^\dagger_{\nu_\mathbf{q}}b_{\nu_{\mathbf{q}}}b^\dagger_{{\nu_2}_{\mathbf{q}_2}}e^{-iq_2d}\ket{\Theta_{q'}e^{-iq'd}}\\
        &= \sum_{\mathclap{\substack{dq\nu j_1 k_1 \\ q_1 q_2 \nu_1 \nu_2}}} e^{i\vec{K}d}\omega_{q\nu}\tilde{\gamma}^{j_1, \nu_1*}_{k_1, q_1}\tilde{\gamma}^{j_1, \nu_2}_{k_1, q_2}e^{-ik_1d}e^{-iq_2d}e^{-iq_1d}e^{-iqd}\Theta^*_{q\nu}\Theta^*_{q_2\nu_2}\Theta_{q_1\nu_1}\Theta_{q\nu}S_d\\
        &+ 2\sum_{\mathclap{\substack{dq\nu j_1k_1\\q_2\nu_2}}}e^{i\vec{K}d}\omega_{q\nu}\tilde{\gamma}^{j_1, \nu*}_{k_1, q}\tilde{\gamma}^{j_1, \nu_2}_{k_1, q_2}e^{-ik_1d}e^{-iq_2d}e^{-iqd}\Theta^*_{q_2\nu_2}\Theta_{q\nu}S_d\\
        &+\sum_{\mathclap{\substack{dq\nu j_1k_1 }}}e^{i\vec{K}d}\omega_{q\nu}\tilde{\gamma}^{j_1, \nu *}_{k_1, q}\tilde{\gamma}^{j_1, \nu}_{k_1, q} e^{-ik_1d}e^{-iqd}S_d\\
        &+\sum_{\mathclap{\substack{dq\nu_1 \nu j_1 k_1 q_1}}}e^{i\vec{K}d}\omega_{q\nu}\tilde{\gamma}^{j_1, \nu_1 *}_{k_1, q_1}\gamma_{k_1, q_1}^{j_1, \nu_1}e^{-ik_1d}e^{-iq_1d}\Theta^*_{q\nu}\Theta_{q\nu}e^{-iqd}S_d.
    \end{aligned}
\end{equation}
Finally, the contribution from the linear electron--phonon coupling term of the Hamiltonian $H_\text{ep}(\Psi)$ can be written as 
\begin{equation}
    \begin{aligned}
        \bra{1}\hat{H}_{\text{ep}}\ket{1} &= \sum_{\mathclap{\substack{dkqk_1k_2\\
        ij\nu j_1 j_2}}} e^{i\vec{K}d}g^{ij}_{\nu}(k,q) \tilde{\zeta}^*_{k_1,j_1}\tilde{\zeta}_{k_2, j_2}\bra{0}\wick{\c1 a_{{j_1}_{{k}_1}} \c1 a^\dagger_{i_{{k}+{q}}} \c1 a_{j_{k}} \c1 a^\dagger_{{j_2}_{{k}_2}}}\ket{0}_{\text{el}}e^{-ik_2 d}\\
        &\bra{\Theta_{q'}}(b^\dagger_{\nu_{-\mathbf{q}}} + b_{\nu_{\mathbf{q}}})\ket{\Theta_{q'\nu}e^{-iq'd}}\\
        &= \sum_{dkqij\nu} e^{i\vec{K}d}g^{ij}_\nu(k,q) \tilde{\zeta}^*_{k+q, i}\tilde{\zeta}_{k, j}e^{-ikd} (\Theta^*_{-q\nu} + \Theta_{q\nu}e^{-iqd})S_d
    \end{aligned}
\end{equation}
\begin{equation}
    \begin{aligned}
        \bra{1}\hat{H}_{\text{ep}}\ket{2} =& \sum_{\mathclap{\substack{dkqk_1k_2q_2 \\ ij\nu j_1 j_2 \nu_2}}} e^{i\vec{K}d}g^{ij}_{\nu}(k,q) \tilde{\zeta}^*_{k_1, j_1}\tilde{\gamma}^{j_2, \nu_2}_{k_2, q_2}\bra{0}\wick{\c1 a_{{j_1}_{{k}_1}} \c1 a^\dagger_{i_{{k}+{q}}} \c1 a_{j_{k}} \c1 a^\dagger_{{j_2}_{{k}_2}}}\ket{0}_{\text{el}}e^{-ik_2d}\\
        &\bra{\Theta_{q'\nu}}(b^\dagger_{\nu_{-\mathbf{q}}} + b_{\nu_\mathbf{q}})b^\dagger_{{\nu_2}_{\mathbf{q}_2}}e^{-iq_2d}\ket{\Theta_{q'\nu}e^{-iq'd}}\\
        =&\sum_{dkqq_2ij\nu\nu_2} e^{i\vec{K}d}g^{ij}_{\nu}(k,q)\tilde{\zeta}^*_{k+q, i}\tilde{\gamma}^{j, \nu_2}_{k, q_2}e^{-ikd}(\Theta^*_{-q\nu} + \Theta_{q\nu}e^{-iqd})\Theta^*_{q_2\nu_2}e^{-iq_2d}S_d\\
        +&\sum_{dkqij\nu}e^{i\vec{K}d}g^{ij}_{\nu}(k,q) \tilde{\zeta}^*_{k+q, i}\tilde{\gamma}^{j, \nu}_{k, q}e^{-ikd}e^{-iqd}S_d
    \end{aligned}
\end{equation}
\begin{equation}
    \begin{aligned}
        \bra{2}\hat{H}_{\text{ep}}\ket{2} &= \sum_{\mathclap{\substack{dkqk_1q_1k_2q_2 \\ij\nu j_1j_2\nu_1\nu_2}}}e^{i\vec{K}d}g^{ij}_{\nu}(k,q)\tilde{\gamma}^{j_1, \nu_1*}_{k_1, q_1}\tilde{\gamma}^{j_2, \nu_2}_{k_2, q_2}\bra{0}\wick{\c1 a_{{j_1}_{{k}_1}} \c1 a^\dagger_{i_{{k}+{q}}} \c1 a_{j_{k}} \c1 a^\dagger_{{j_2}_{{k}_2}} }\ket{0}_{\text{el}}e^{-ik_2d}\\
        &\bra{\Theta_{q'\nu}}b_{{\nu_1}_{\mathbf{q}_1}}(b^\dagger_{\nu_{-\mathbf{q}}} + b_{\nu_{\mathbf{q}}})b^\dagger_{{\nu_2}_{\mathbf{q}_2}}e^{-iq_2d}\ket{\Theta_{q'\nu}e^{-iq'd}}\\
        &= \sum_{\mathclap{\substack{dkq q_1 q_2 \\ ij\nu \nu_1\nu_2}}}e^{i\vec{K}d}g^{ij}_\nu(k,q) \tilde{\gamma}^{i, \nu_1*}_{k+q, q_1}\tilde{\gamma}^{j, \nu_2}_{k, q_2}e^{-ikd}e^{-iq_2d}e^{-iq_1d} (\Theta^*_{-q\nu} + \Theta_{q\nu}e^{-iqd})\Theta^*_{q_2\nu_2}\Theta_{q_1\nu_1}S_d\\
        &+\sum_{\mathclap{\substack{dkqq_2\\ij\nu \nu_2}}}e^{i\vec{K}d}g^{ij}_{\nu}(k,q)\tilde{\gamma}^{i, \nu*}_{k+q, -q}\tilde{\gamma}^{j, \nu_2}_{k, q_2}e^{-ikd}e^{-iq_2d}\Theta^*_{q_2\nu_2}S_d\\
        &+\sum_{\mathclap{\substack{dkqq_2\\ij\nu \nu_2}}}e^{i\vec{K}d}g^{ij}_{\nu}(k,q)\tilde{\gamma}^{i, \nu_2*}_{k+q, q_2}\tilde{\gamma}^{j, \nu_2}_{k, q_2}e^{-ikd}e^{-iq_2d}(\Theta^*_{-q\nu} + \Theta_{q\nu}e^{-iqd})S_d\\
        &+\sum_{\mathclap{\substack{dkqq_1\\ ij\nu \nu_1}}}e^{i\vec{K}d}g^{ij}_{\nu}(k,q)\tilde{\gamma}^{i, \nu_1*}_{k+q, q_1}\tilde{\gamma}^{j, \nu}_{k,q}e^{-ikd}e^{-iqd}e^{-iq_1d}\Theta_{q_1\nu_1}S_d.
    \end{aligned}
\end{equation}
Collecting the terms as $H(\Psi) = H_\text{el}(\Psi) + H_\text{ph}(\Psi) + H_{\text{ep}}(\Psi)$, we may then compute the first-order variational energy as $E_{\text{dCC-1-S1}} = \frac{H(\Psi)}{O(\Psi)}$.
\section{\label{app:secondorder}Second-order Energy Evaluation}
In this appendix we detail the computation of second-order energy expectation values. For convenience, we work in real-space assuming a single electronic and phonon band. When evaluating the wavefunction overlap, we encounter terms of the form
\begin{equation}
\label{eq:squeeze_exp}
    \bra{\boldsymbol{\Theta}}e^{\frac{1}{2}\sum_{ij}\beta^*_{ij}b_ib_j}e^{\frac{1}{2}\sum_{ij}\beta_{i-d, j-d}b^\dagger_i b^\dagger_j}\ket{\boldsymbol \Theta_{-d}}
\end{equation}
where $\ket{\boldsymbol \Theta_{-d}}$ denotes an unnormalized coherent state with coherent state shifts $\boldsymbol \Theta$ permuted by $d$ sites as imposed by the momentum projection.
\begin{equation}
    \ket{\boldsymbol \Theta_{-d}}:= \exp(\sum_i \Theta_{i-d}b^\dagger_i)\ket{0}
\end{equation}
Inserting a coherent state resolution of identity over the independent modes, we have
\begin{equation}
\begin{aligned}
\label{eq:overlapsqueeze}
    &\int \frac{d\bar{\boldsymbol \alpha}d\boldsymbol \alpha}{(2\pi i)^N} e^{-\sum_j \bar{\alpha}_j \alpha_j}\bra{\boldsymbol \Theta}e^{\frac{1}{2}\sum_{ij}\beta^*_{ij}b_i b_j}\ket{\boldsymbol \alpha}\bra{\boldsymbol \alpha}e^{\frac{1}{2}\sum_{ij}\beta_{i-d, j-d} b^\dagger_i b^\dagger_j}\ket{\boldsymbol \Theta_{-d}}\\
    =&\int \frac{d\bar{\boldsymbol \alpha}d \boldsymbol\alpha}{(2\pi i)^N} \exp(-\sum_j \bar{\alpha_j}\alpha_j + \frac{1}{2}\sum_{ij}\beta^*_{ij}\alpha_i\alpha_j + \sum_j \Theta^*_j\alpha_j + \sum_j \bar{\alpha}_j\Theta_{j-d} + \frac{1}{2}\sum_{ij}\beta_{i-d,j-d} \bar{\alpha}_i\bar{\alpha}_j)
\end{aligned}
\end{equation}
which may be written in matrix form by defining
\begin{equation}
    Z = \begin{pmatrix}
        \boldsymbol{\alpha} \\ \boldsymbol{\bar{\alpha}} 
    \end{pmatrix} \in \mathbb{C}^{2N}, \quad 
    Q = \begin{pmatrix}
        \boldsymbol \beta^* & -\mathbb{I}\\
        -\mathbb{I} & \tilde{\boldsymbol \beta}
    \end{pmatrix} \in \mathcal{M}_{2N\times 2N}(\mathbb{C}), \quad 
    R =
\begin{pmatrix}
    \boldsymbol{\Theta}^* \\
    \boldsymbol{\Theta}_{-d}
\end{pmatrix},
\end{equation}
with $\tilde{\boldsymbol \beta}_{ij} := \beta_{i-d, j-d}$ and $(\boldsymbol{\Theta}_{-d})_j = \Theta_{j-d}$. We can then recast \cref{eq:overlapsqueeze} as a Gaussian integral
\begin{equation}
    \int \frac{d\bar{\boldsymbol\alpha}d\boldsymbol\alpha}{(2\pi i)^N} \exp(\frac{1}{2}Z^TQZ + R^TZ),
\end{equation}
which upon transforming to real variables with
\begin{equation}
    Z=TX = \begin{pmatrix}
        \mathbb{I} & i\mathbb{I}\\
        \mathbb{I} & -i\mathbb{I}
    \end{pmatrix}\begin{pmatrix}
        x\\y
    \end{pmatrix}, \quad M = -T^TQT, \quad J= T^TR
\end{equation}
can be written as
\begin{equation}
    \frac{1}{\pi ^N}\int dX\; \exp(-\frac{1}{2}X^TMX + J^TX).
\end{equation}
This can then be evaluated using standard techniques \cite{OrlandNegele}. Making the change of variables $Y = X-M^{-1}J$ and completing the square, we have
\begin{equation}
    \frac{1}{\pi ^N}\int dY \exp(-\frac{1}{2}Y^TMY + \frac{1}{2}J^TM^{-1}J), 
\end{equation} which can be solved to obtain
\begin{equation}
\label{eq:generating}
    I(J):=\frac{2^N}{\sqrt{\det (M)}}e^{\frac{1}{2}J^TM^{-1}J}.
\end{equation}
Energy expectation values involve bosonic strings of $b^\dagger\ldots b$ inserted between the squeezing operators in \cref{eq:squeeze_exp}, which may be evaluated by taking appropriate derivatives of the generating function $I(J)$ with respect to the source vector $J$. Noting that insertions of $b^\dagger_{k+d} $ are transformed into factors of $\bar{\alpha}_{k+d}$ following an application of the resolution of identity in \cref{eq:squeeze_exp}, and likewise $b_j $ to $\alpha_j$, we may evaluate expectation values of normal ordered strings of bosonic operators by applying the rule
\begin{equation}
    b_j \rightarrow v_j^T\partial_J, \quad b^\dagger_{k+d} \rightarrow v^{(\dagger)}_{k+d} \partial_J
\end{equation}
with 
\begin{equation}
    v^{(\dagger)}_{k+d}:= \begin{pmatrix}
        e_{k+d}\\ -ie_{k+d}
    \end{pmatrix}^T \in \mathbb{C}^{2N},\quad v_k^T:= \begin{pmatrix}
        e_k\\ ie_k
    \end{pmatrix}^T \in \mathbb{C}^{2N},
\end{equation}
where we have let $e_i = (0,..,1_i, ..0)$. Expectation values of bosonic strings $\langle b^\dagger \ldots b\rangle$ may then be evaluated by taking the appropriate derivatives of the generating function \cref{eq:generating}.
\section{Fr\"ohlich model}\label{app:Frohlich}
In this appendix, we detail the discretization procedure, reference states employed, and the continuum-extrapolation protocol underlying the Fr\"ohlich results reported in \cref{sec:frohlich}. The imposition of a periodic finite system volume $\Omega$ in the Fr\"ohlich coupling $g_{\text{F}}(\mathbf{q})$ restricts the allowed momenta to the set $\frac{2\pi}{L} \mathbb{Z}^3$, where $\Omega = L^3$.

The discretization of the real-space volume element with a grid of spacing $a = \frac{L}{N}$ further restricts the set of allowed momenta to 
\begin{equation}
    \mathbf{k}  = \frac{2\pi}{L}\mathbf{n},\quad n_i \in \left\{-\frac{N}{2}, \dots , \frac{N}{2}-1\right\}
\end{equation}
so that each Cartesian component is bounded by the momentum cutoff  $k_\text{max} = \frac{\pi}{a}$. Momentum addition $\mathbf{k} + \mathbf{q}$ is folded back onto the grid modulo $\frac{2\pi}{a}$. The discretized model thus has two parameters with disjoint physical roles: the box size $L=Na$ controls the infrared behavior, while the cutoff $k_\text{max} = \frac{\pi}{a}$ controls the ultraviolet. The continuum Fr\"ohlich model is then recovered in the ordered double limit $L\rightarrow \infty$ at fixed $a$, followed by $a\rightarrow 0$. Such a procedure is necessary as the limit $a\rightarrow 0$ at fixed $N$ results in an increasingly small volume and does not converge.

The Fr\"ohlich coupling $g_\mathbf{q}$ at $\mathbf{q}=0$ is formally singular and requires careful handling on a discrete grid. Though this element is conventionally excluded, that exclusion leads to a slowly converging finite-size error. As the $\mathbf{q}=0$ phonon mode couples exactly to the total electron number $\sum_\mathbf{k} a^\dagger_\mathbf{k}a_\mathbf{k}$, which is unity for the single electron considered here, its Hamiltonian contribution is that of a displaced harmonic oscillator, which decouples exactly and contributes $-g_{\mathbf{0}}^2$ to the ground-state energy for any value assigned to the coupling $g_{\mathbf{0}}$ at that point. On a finite grid, the singular grid point represents a momentum space cell of volume $\Omega_0 = \frac{(2\pi)^3}{\Omega}$. Though the Fr\"ohlich coupling element is singular at this point, its corresponding contribution to the energy is finite at finite $L$ because the $\frac{1}{q^2}$ singularity in $|g_\mathbf{q}|^2$ is integrable in three dimensions due to its cancellation with the three-dimensional volume element. Viewing the discrete grid as a midpoint quadrature approximation of the continuum limit, we therefore replace the singular $\mathbf{q}=0$ contribution with an effective central cell coupling chosen to approximate the integrated value of $|g_\mathbf{q}|^2$ over that cell, analogous to standard treatments of integrable long-wavelength singularities in periodic calculations \cite{Chiesa2006, Fraser1996}. We approximate the integration volume with an equivalent volume sphere with radius $q_c = \left(\frac{3\Omega_0}{4\pi}\right)^{1/3}$, and obtain the analytic average in the central cell of
\begin{equation}
\begin{aligned}
    \langle g_\mathbf{q}^2 \rangle_0 &= \frac{1}{\Omega_0}\int_0^{q_c}\int_0^\pi\int_0^{2\pi} \frac{2^{3/2}\pi \alpha }{\Omega}\frac{1}{q^2}q^2 \sin(\theta)dqd\theta d\phi \\
    &= \frac{\sqrt{2} \alpha }{\pi}q_c\\
    &= {\sqrt{2}\alpha}\left(\frac{6}{\pi}\right)^{1/3}\frac{1}{L}.
\end{aligned}
\end{equation}
We thus see that excluding the $\mathbf{q}=0$ element introduces a finite-size error which decays as $O(L^{-1})$, and replace the singular coupling with $\sqrt{\langle g_{\mathbf{q}}^2\rangle_0}$ analogous to procedures employed in previous studies \cite{VasilchenkoFrohlich, Miglio2020}. As this mode couples only to the total electron number and thus decouples completely, the error incurred by approximating the volume with a spherical cell vanishes identically in the reported corrected values, in which this error cancels with the subtracted auxiliary quantity sharing this approximation.
\subsection{Reference choice}\label{app:frohlich-ref}
The dCC calculations of this appendix employ the discrete machinery of the main text and require only a suitable choice of reference state. We use a coherent-state product reference in which the electron occupies an orbital drawn from a one-parameter family, with coherent-state shifts determined self-consistently from the electronic density. For $\alpha \leq 9$, we use a Gaussian orbital of width $\sigma$, while for  $\alpha \geq 10$ we use the scaled Pekar orbital 
\begin{equation}
    \varphi_w(\mathbf{r}) = (\alpha w)^{3/2} \varphi_\text{P}(\alpha w\mathbf{r})
\end{equation}
where $\varphi_\text{P}(\mathbf{r})$ minimizes the strong-coupling Pekar functional \cite{Pekar1954, Miyake1975} at unit coupling and is obtained numerically. As the Pekar orbital is only exact in the adiabatic continuum limit, the scaling parameter $w$ is retained to absorb the leading finite-grid and finite-$\alpha$ corrections to the orbital profile. In particular, the scale parameter ($\sigma$ or $w$) is chosen to minimize the closed-form discrete grid energy with the constraint that the kinetic energy of the discrete orbital match its continuum counterpart to a fixed tolerance. This guards against artificial overbinding of the polaron for sufficiently localized states caused by the saturation of kinetic energy at the fixed momentum cutoff scale and the continual growth of the Fr\"ohlich binding energy. Importantly, the continuum reference energy remains analytic as a homogeneous function of the scaling parameter $w$, which allows the exact finite-size and finite-cutoff error of the discrete reference to be determined in closed form. As an additional safeguard, the momentum-space grid weight of the optimized dCC electronic field is verified to vanish at the grid edge, which excludes the artificially overbound lattice states, whose weight there remains $O(1)$. 
\subsection{Finite-size correction and continuum convergence}\label{app:frohlich-protocol}
Each converged dCC energy is corrected by the exact discretization error of an auxiliary energy, whose continuum limit is known in closed form. We apply three such auxiliary quantities. 

With reference orbital coefficients $\phi_\mathbf{k}$ and the grid-density form
\begin{equation}
    n_\mathbf{q} = \sum_\mathbf{k} \phi^*_{\mathbf{k}+\mathbf{q}}\phi_\mathbf{k},
\end{equation}
the minimized product state reference energy on the grid takes the closed form
\begin{equation}
\label{eq:ref_e_grid}
    E^\text{grid}_\text{ref} = \sum_\mathbf{k} \frac{|\mathbf{k}|^2}{2}|\phi_\mathbf{k}|^2 - \sum_\mathbf{q}g_\mathbf{q}^2n_\mathbf{q}^2,
\end{equation}
while its continuum limit remains analytic for every reference we employ. We then have $E^\infty(\sigma) = \frac{3}{4\sigma^2} - \frac{\alpha }{\sigma\sqrt{\pi}}$ for the Gaussian family, and $E^\infty(\alpha, w) = \alpha^2(T_1w^2 - C_1 w)$ for the scaled Pekar orbital, where
\begin{equation}
    T_1 = \frac{1}{2}\int |\nabla \varphi_\text{P}|^2 = 0.108513,\quad \text{and } C_1 = 2T_1
\end{equation}
by the virial condition. The third auxiliary is second-order perturbation theory, which is not used as a reference, but has grid value $E^\text{grid}_\text{PT2}=-\sum_\mathbf{q} \frac{g_\mathbf{q}^2}{1+|\mathbf{q}|^2/2}$ and continuum limit $E^\infty_\text{PT2} = -\alpha$. As the dCC ansatz is complete in the one-phonon excitation sector, its optimized amplitudes at weak coupling assume the exact perturbative values at every retained momentum, and the discretization error of the dCC energy thus coincides with that of second-order perturbation theory to leading order in the coupling.

We then report
\begin{equation}
\label{eq:cv}
    E_\text{corrected} = E_\text{dCC} + \left[E^\infty_{\text{aux}} - E^\text{grid}_{\text{aux}}\right]
\end{equation}
where the bracketed expression is the exact finite-size and finite-cutoff correction of the auxiliary and independent of the dCC optimization and benchmark quantities, so that \cref{eq:cv} removes from the dCC energy the part of its discretization error which the auxiliary shares. Writing $\delta X = X^\text{grid} - X^\infty$, we have
\begin{equation}
    E_\text{corrected} - E^\infty_{\text{dCC}} = \delta(E_\text{dCC}- E_\text{aux})
\end{equation}
which is small precisely when the auxiliary quantity shares the same physical structure as our ansatz, which is perturbative at weak couplings and adiabatic at strong couplings. The auxiliary is selected by two \textit{a priori} rules. First, the auxiliary must carry a discretization error of definite sign, as the dCC error it corrects approaches from one side. The perturbative auxiliary loses this property for $\alpha \gtrsim 8$, where its infrared and ultraviolet errors enter with opposing signs and become comparable within our range of box sizes, so that the bracket passes through zero mid sequence with respect to the grid size. Second, among the qualified auxiliaries, we select the one whose corrected sequence is flattest across the largest grids employed, with agreement between independent auxiliaries serving as a consistency check. The corrected energy is thus a finite-size corrected estimate with a controlled residual and is not guaranteed to be variational, while the raw energy of each grid remains a strict upper bound of the discretized model. The auxiliary dependence itself diminishes systematically with cutoff, with the perturbative and Gaussian corrections at $\alpha = 6$ approaching one another from opposite sides as $k_\text{max}$ grows, so that the selection matters least where the extrapolation is read.

The two continuum limits are treated in separate ways suitable for their measured residuals. The box limit is read directly at its demonstrated plateau over the largest grids, where the final box size changes the energy by less than $6\times 10^{-3}$ per unit coupling, and is not extrapolated. The momentum cutoff is extrapolated with a power law form, and the box-converged plateaus drawn from $k_\text{max}/\alpha = \{1.4, 1.75, 2, 2.5, 3\}$, subject to the infrared floor $k_\text{max}\geq 4$ are extrapolated as
\begin{equation}
    E(k_\text{max}) = E_\infty +c k_\text{max}^{-p}
\end{equation}
with a single exponent shared across all couplings and per-$\alpha$ coefficients $E_\infty, c$. An exponent fit per $\alpha$ would render each three-point fit exactly determined, while the shared exponent leaves one validating degree of freedom per coupling and yields $p\simeq 2.5$. Reported uncertainties vary the exponent over $p\in [1.6, 3.6]$, which is determined by considering the range of $p$ which produce a residual within two times that of the optimized $p$. The power law form is validated at $\alpha =10$, where a fourth cutoff $k_\text{max} = 3\alpha$ was computed after fitting and agrees with the fit-prediction to $3\times 10^{-4}$ per unit coupling. We use the same data to isolate the pure cutoff error by comparing simulations done with identical box sizes.

The D2 (Landau--Pekar) product ansatz energy was separately computed directly on discretized grids at a fixed ultraviolet cutoff of $k_\text{max}/\alpha=4$ which was numerically confirmed to leave a negligible residual contribution to the ground-state energy. The resulting energies obtained on different box sizes were then extrapolated with respect to the box length $L$. As the D2 ansatz reduces to the Pekar functional in the continuum limit, which has a known global minimum of $E_\infty/\alpha^2 = -0.108513$ \cite{Pekar1954, Miyake1975}, this provides a check against a known limit.
\subsection{Ground-state energies and benchmark comparison}
\Cref{tab:frohlich} collects the dCC ground-state energies across the coupling range together with digitized diagrammatic Monte Carlo values \cite{Hahn2018}. A comparison of the ground-state energies obtained via different methods is shown in \cref{fig:frohlich_ground}. The raw energy of the largest grid run at a reference cutoff $k_\text{max} = 2\alpha$ is reported unmodified, and serves as a variational upper bound to the corresponding discretized model. The continuum estimate $E_\infty$ follows the protocol of Appendix~\ref{app:frohlich-protocol}, and its quoted uncertainty represents half the spread in energy obtained by refitting each coupling with the exponent range endpoints $\{1.6, 3.6\}$. Only $\alpha=1$ is reported with no extrapolation. At $\alpha = 1$, all cutoff ratios fall below the floor $k_\text{max}=4$, and we thus report the converged value at $k_\text{max}=4$ directly. 
\begin{table}[h]
  \begin{tabular}{ccccc}
       $\alpha$ & $E_\text{raw}$ & $E_\infty$ & $E_\text{DiagMC}$ & Err.\\ \hline
       1  & $-0.8195^{\,b}$  & $-1.01657^{\,b}$ & $-1.01662$  & $+0.00\%$ \\
       2  & $-1.6772$  & $-2.06882(66)$& $-2.06957$  & $+0.04\%$ \\
       3  & $-2.7346$ & $-3.1582(20)$& $-3.16829$  & $+0.32\%$ \\
       4  & $-3.8391$  & $-4.2822(62)$ & $-4.32490$  & $+0.99\%$ \\
       5  & $-5.0032$  & $-5.474(11)$ & $-5.55297$  & $+1.43\%$ \\
       6  & $-6.2647$  & $-6.732(20)$& $-6.86647$  & $+1.96\%$ \\
       7  & $-7.6146$  & $-8.100(24)$& $-8.31039$  & $+2.54\%$ \\
       8  & $-9.0672$  & $-9.670(74)$& $-9.92206$  & $+2.54\%$ \\
       9  & $-10.7426$ & $-11.466(78)$& $-11.72535$ & $+2.21\%$ \\
      10  & $-12.6375$ & $-13.58(11)$& $-13.78200$ & $+1.45\%$ \\
      11  & $-14.7391$ & $-15.88(16)$& $-16.06600$ & $+1.17\%$ \\
      12  & $-17.0398$ & $-18.40(19)$& $-18.59430$ & $+1.05\%$ \\
      13  & $-19.5202$ & $-21.14(23)$& $-21.24340$ & $+0.50\%$ \\
      14  & $-22.2109$ & $-24.09(27)$& $-24.11510$ & $+0.10\%$ \\
      15  & $-25.0818$ & $-27.22(29)$& $-27.26290$ & $+0.15\%$ \\
  \end{tabular}
\caption{Fr\"ohlich ground-state energies in units of $\hbar\omega_\text{LO} = 1$. $E_\text{raw}$ is the raw dCC energy obtained from the largest grid at the momentum cutoff $k_\text{max}=2\alpha$ used at a particular coupling without any auxiliary correction. $E_\infty$ is the continuum estimate obtained from extrapolation with respect to momentum cutoff, with the approximate uncertainty in the last decimal shown in parentheses. The superscript $b$ denotes that the value is reported without extrapolation using the cutoff $k_\text{max}=4$, and applies only to the $\alpha=1$ value. The displayed error is computed from $(E_\infty - E_\text{DiagMC})/|E_\text{DiagMC}|$.}
\label{tab:frohlich}
\end{table}

At weak couplings, the corrected energies are in excellent quantitative agreement with the benchmark where the perturbative matching of the one-phonon sector operates. The deviation achieves its maximum in the crossover regime of approximately $2.54\%$ at $\alpha =8$ where neither the perturbative nor adiabatic description is sufficiently accurate. This error progressively decreases at stronger couplings where the scaled Pekar reference becomes increasingly accurate. Though each extrapolated energy lies above the benchmark, the lower edge of the exponent band extends between $0.1\%$ and $1.4\%$ below the digitized values across $\alpha = 11-15$, with the extrapolated limit and rest of the band lying above.

\section{Additional numerical results}\label{app:addl_results}
\subsection{Dispersive-phonon band structure}
We augment the 1D Holstein model with a phonon hopping $t_{\text{ph}}\sum_{i}b^\dagger_{i+1}b_i + \text{h.c.}$, which yields the phonon dispersion $\omega(q) = \omega_0 + 2t_{\text{ph}}\cos(q)$ with $\omega_0$ the Einstein frequency.
\begin{figure}[tb]
\centering
\includegraphics[width=0.5\linewidth]{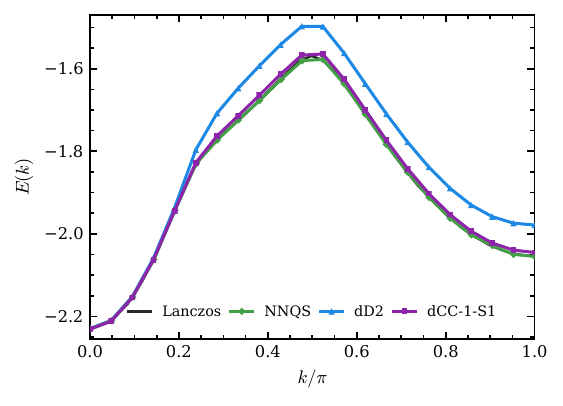}
\caption{\label{fig:1d_disp} \textbf{Polaron band structure of the one-dimensional Holstein model with dispersive phonons.} Band structure of the 42-site Holstein model with dispersive phonons ($t_{\text{ph}} = 0.4$) compared with the dD2 ansatz, NNQS \cite{mahajan24}, and Lanczos diagonalization \cite{Bonca_disp}.}
\end{figure}
\Cref{fig:1d_disp} compares the resulting polaron band structure with numerically exact Lanczos benchmarks obtained directly in the thermodynamic limit \cite{Bonca_disp} and NNQS calculations on a 42-site lattice \cite{mahajan24}. dD2 is highly accurate at small polaron momenta for this coupling type but incurs a large error near the Brillouin-zone edge, which the first-order correlated excitations in dCC-1-S1 correct.

\subsection{Real-space observables}
\begin{figure}[tb]
    \centering
    \includegraphics[width=0.5\linewidth]{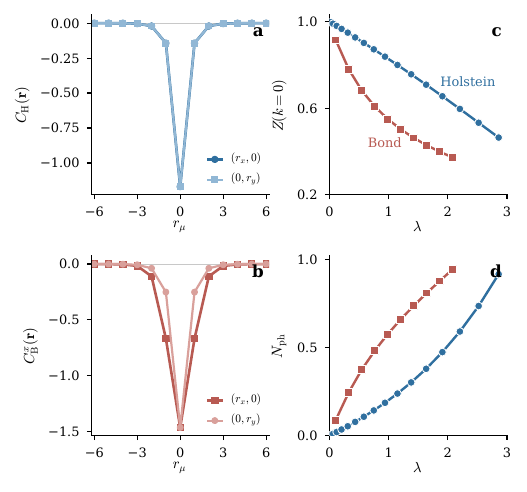}
    \caption{\textbf{Two-dimensional Holstein and BSSH observables from dCC-1-S1.} We consider the anti-adiabatic regime $\omega =3, t= 1$, computed on a $20\times20$ lattice. (a) Holstein density-displacement correlation function $C_\text{H}(\mathbf{r})$ computed at $\lambda \approx 1.63$. (b) bond-displacement correlation function $C_\text{B}^x(\mathbf{r})$ computed at $\lambda \approx 1.64$. Panels (a,b) show one-dimensional cuts along $\mathbf{r}= (r_x,0)$ and $\mathbf{r} = (0, r_y)$. (c) Quasiparticle residue $Z_{k=0}$ and (d) total mean phonon number as functions of $\lambda = g^2/2\omega t.$
    }
    \label{fig:observables}
\end{figure}

For the Holstein model, the density-displacement correlation
\begin{equation}
    C_\text{H}(\mathbf{r}) := \langle  \sum_i \hat{n}_i \hat{X}_{i+\mathbf{r}}\rangle,
\end{equation}
with $\hat{X}_i$ the displacement operator $b^\dagger_i + b_i$, measures the lattice deformation around the carrier, while for the BSSH model the corresponding bond correlation function in the $x$-direction,
\begin{equation}
    C_\text{B}^x(\mathbf{r}) : = \langle \sum_i (a^\dagger_{i+\hat{x}}a_i + a^\dagger_{i} a_{i+\hat{x}})\hat{X}^x_{i+\mathbf{r}}\rangle,
\end{equation}
characterizes the degree to which the electron--phonon interaction modulates the hopping along a given direction. We additionally compute the quasiparticle weight $Z_k= |\bra{0}a_k \ket{\Psi}|^2/\braket{\Psi}{\Psi}$ for the ground state $\ket{\Psi}$ and the average phonon number $N_{\text{ph}}= \langle \sum_i b^\dagger_i b_i\rangle$ as functions of coupling strength in the anti-adiabatic regime $\omega \gg t$. The results are shown in \cref{fig:observables} and discussed in the main text.

\section{Tangent-space structure}{\label{app:tangent}}
In the following, we will detail the tangent-space basis states included within the first- and second-order levels of excited state theory, working in a real-space, single-band formalism suitable for the model systems considered for simplicity. Throughout this appendix, both the reference wavefunction and its tangent vectors use unnormalized Bargmann coherent states for simplicity. In this setting, the first-order wavefunction may be expressed as 
\begin{equation}
\begin{aligned}
    \ket{\Psi_{\text{dCC-1-S1}}} &= \hat{\Xi} e^{\hat{T}}\ket{\Psi_{\text{D}2}}\\
    &= \hat{\Xi}e^{\hat{T}}\left(\sum_l \psi_l a^\dagger_l \ket{0}_\text{el} \otimes \exp(\sum_m \alpha_m b^\dagger_m)\ket{0}_\text{ph}\right) \\
    &=\hat{\Xi}\left(\sum_l \zeta_l a^\dagger_l \ket{0}_{\text{el}}\otimes \ket{\boldsymbol\Theta} + \sum_{lm} \gamma_{l,m} a^\dagger_l \ket{0}_{\text{el}}\otimes b^\dagger_m\ket{\boldsymbol\Theta} \right)
\end{aligned}
\end{equation}
where the coherent state shifts $\boldsymbol \Theta := \boldsymbol \alpha + \boldsymbol \beta$ are composed of a fixed part $\boldsymbol \alpha$ obtained from the reference D2 calculation and a component $\boldsymbol \beta : = (t_1, t_2, \cdots , t_N)$ introduced through the coupled cluster operator $e^{\hat{T}_\text{ph}} = \exp(\sum_l t_l b^\dagger_l)$ . We have additionally defined the single-band specializations of $\zeta_l$ and $\gamma_{l,m}$ defined in terms of the real-space coupled-cluster amplitudes and D2 electronic parameters ($\{\psi\}$)
\begin{equation}
    \zeta_l : = \psi_l + \sum_a t_0^a \psi_{l,a}
\end{equation}
\begin{equation}
    \gamma_{l,m} := \sum_a t_{0,m}^a \psi_{l,a}
\end{equation}
where $0$ denotes the single occupied electronic orbital and $a$ labeling the virtual orbitals ($\psi_{1,a}, ... \psi_{N,a}$) defined to form a complete orthonormal set. 

As discussed in the main text, the tangent states are obtained by taking derivatives of the ground-state wavefunction with respect to the variational parameters. At the first-order level of theory for the ground state, there are three types of parameters describing electronic, phonon, and coupled electron--phonon degrees of freedom. This leads to the three types of tangent states
\begin{equation}
\begin{aligned}
    \ket{1_\text{e}}&:= \frac{\partial }{\partial t_{0}^a}\ket{\Psi_{\text{dCC-1-S1}}} = \hat{\Xi} \left(\sum_{\vec{l}} \psi_{\vec{l},a} a^\dagger_l \ket{0}_{\text{el}} \otimes \ket{\boldsymbol\Theta}\right)\\
    \ket{2_\text{p}} &:= \frac{\partial }{\partial t_z}\ket{\Psi_{\text{dCC-1-S1}}} = \hat{\Xi} \left(\sum_{\vec{l}} \zeta_{\vec{l}}a^\dagger_{\vec{l}}\ket{0}_{\text{el}} \otimes b^\dagger_{\vec{z}}\ket{\boldsymbol \Theta}+\sum_{\vec{l}\vec{k}} \gamma_{\vec{l},\vec{k}}a^\dagger_{\vec{l}} \ket{0}_{\text{el}} \otimes b^\dagger_{\vec{k}}b^\dagger_{\vec{z}}\ket{\boldsymbol \Theta}\right)\\
    \ket{3_\text{ep}} &:= \frac{\partial}{\partial t^a_{0,z}}\ket{\Psi_{\text{dCC-1-S1}}} = \hat{\Xi}\left(\sum_{\vec{l}} \psi_{\vec{l}, a} a^\dagger_{\vec{l}} \ket{0}_{\text{el}} \otimes b^\dagger_{\vec{z}}\ket{\boldsymbol \Theta}\right).
\end{aligned}
\end{equation}
We see from the above that these states describe single electronic excitations and one phonon corrections to the ground state at first order. At the second order of excited state theory, the additional second-derivative states
\begin{equation}
    \begin{aligned}
        \ket{4_{\text{pp}}} &:= \frac{\partial^2}{\partial t_z \partial t_x}\ket{\Psi_{\text{dCC-1-S1}}} = \hat{\Xi}\left(\sum_{\vec{l}}\zeta_{\vec{l}}a^\dagger_{\vec{l}}\ket{0}_{\text{el}}\otimes b^\dagger_{x}b^\dagger_{z}\ket{\boldsymbol\Theta}+\sum_{\vec{l}\vec{k}}\gamma_{\vec{l},\vec{k}}a^\dagger_{\vec{l}}\ket{0}_{\text{el}}\otimes b^\dagger_{x}b^\dagger_{z}b^\dagger_{\vec{k}}\ket{\boldsymbol\Theta}\right) \\
        \ket{5_{\text{ep}}} &:= \frac{\partial^2}{\partial t^a_{0,z}\partial t_x} \ket{\Psi_{\text{dCC-1-S1}}} = \hat{\Xi} \left(\sum_{\vec{l}}\psi_{\vec{l},a}a^\dagger_{\vec{l}}\ket{0}_{\text{el}} \otimes b^\dagger_{\vec{z}}b^\dagger_{\vec{x}}\ket{\boldsymbol \Theta}\right)
    \end{aligned}
\end{equation}
which incorporate two-phonon corrections on top of the ground state. We note that mixed-partial derivatives of electron and phonon degrees of freedom yield states identical to the tangent directions corresponding to first derivatives of the coupled electron--phonon parameters, and second-order electronic derivatives vanish, so that these directions do not contribute to the second-order manifold. All subsequent required matrix elements involving these states may be computed in an analogous way to the first-order ground-state theory.

\begin{figure}[tb]
    \centering
    \includegraphics[width=0.55\linewidth]{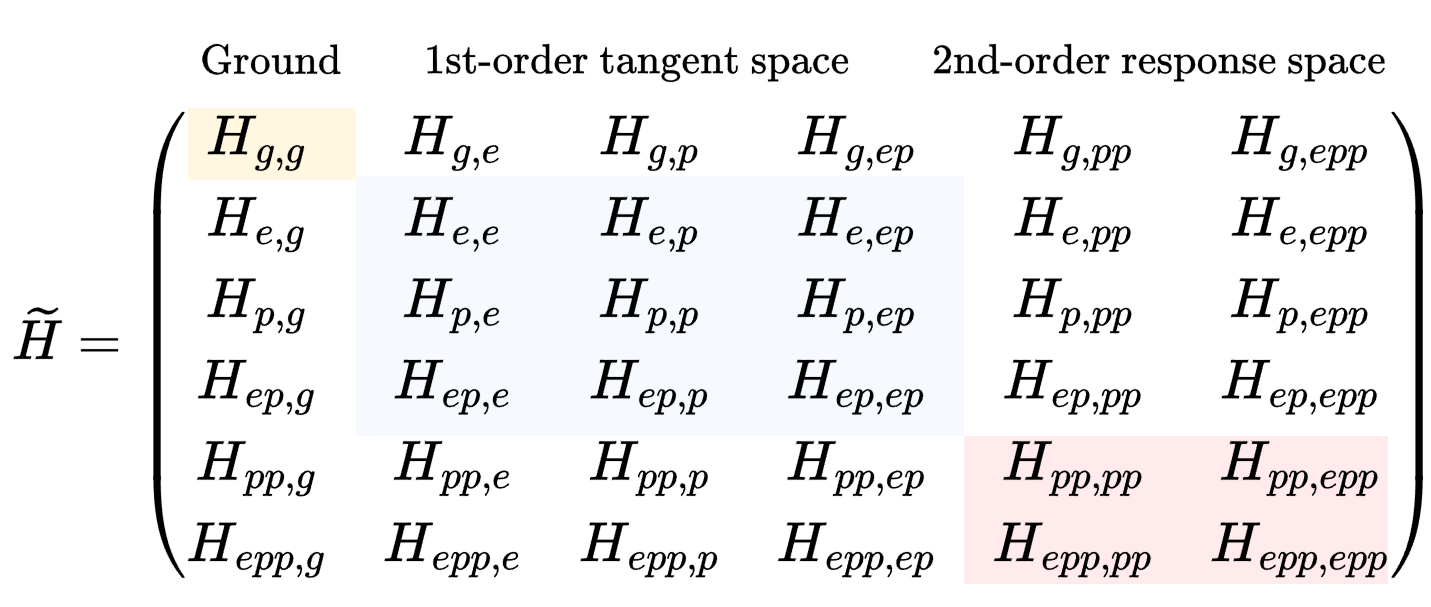}
    \caption{\textbf{Block structure of the projected response matrices at first and second order in the linear-response hierarchy.} The shaded regions denote the ground, first-order (dCC-1-S1-(1,1)), and second-order  (dCC-1-S1-(2,2)) truncated manifolds respectively.
    }
    \label{fig:EOM}
\end{figure}
The projected Hamiltonian and overlap matrices are organized in blocks corresponding to the electronic, phononic, coupled electron--phonon, two-phonon, and coupled electron-two-phonon ($\text{e, p, ep, pp, epp}$) sectors, as illustrated in \cref{fig:EOM}. In the present work, we evaluate and store these matrices explicitly and solve the resulting generalized eigenvalue problem directly, which is sufficient for the benchmark systems considered here. For larger-dimensional problems, the same formalism can be paired with iterative eigensolvers \cite{davidson1975} and sparse solvers \cite{Lehoucq1998} by using matrix-vector products without storing the full matrix, with the associated costs summarized in \cref{tab:response_scaling}.

\section{\label{app:finite_T} Finite-temperature matrix elements}
In this appendix we sketch the evaluation of necessary matrix elements needed in the first-order finite-temperature implementation. To begin, the explicit time- and temperature-dependence from the finite-temperature trace and Heisenberg time evolution may be absorbed into the coherent-state basis as
\begin{equation}
\begin{aligned}
    \ket{\boldsymbol\sigma(\beta,t)} &:=e^{\sum_{mn} (-\beta - it)\omega_{mn}b^\dagger_m b_n}\ket{\boldsymbol \alpha}\\
    &= \exp(\left[e^{-(\beta + it)\boldsymbol \omega} \cdot \boldsymbol{\alpha}\right]^T \mathbf{b}^\dagger)\ket{0}
\end{aligned}
\end{equation}
using a BCH expansion, where $\boldsymbol \omega $ is the matrix describing the phonon dispersion, $\boldsymbol \alpha$ the vector of coherent state shifts per site, and $\mathbf{b}^\dagger$ a vector containing creation operators per site $(b^\dagger_1, \cdots ,b^\dagger_N)$. As seen in the preceding equation, the time- and temperature-dependence are accounted for by introducing a time and temperature dependence into the coherent state shifts of the tracing coherent state basis. 

When evaluating the finite-temperature Green's function, we encounter integrals of matrix elements of the form
\begin{equation}
\begin{aligned}
\label{eq: ft_int}
    &\int \frac{d\boldsymbol \alpha d \bar{\boldsymbol \alpha}}{(2\pi i)^N} e^{-\sum_z \bar{\alpha}_z \alpha_z}\bra{\boldsymbol\sigma(\beta, t)}\ket{\boldsymbol \Theta_{-d}}\bra{\boldsymbol \Theta_{-d'}}\ket{\boldsymbol{\alpha}}\\
    =&\int \frac{d\boldsymbol \alpha d \bar{\boldsymbol \alpha}}{(2\pi i)^N} \exp(-\sum_z \bar{\alpha}_z\alpha_z + \sum_z \Theta^*_{z-d'}\alpha_z + \sum_z \left[e^{-(\beta + it)\boldsymbol \omega} \cdot \boldsymbol \alpha\right]^*_z \Theta_{z-d}).
\end{aligned}
\end{equation}
Similarly to the second-order energy evaluation, this may be written in matrix-vector notation as 
\begin{equation}
    \int \frac{d\boldsymbol \alpha d \bar{\boldsymbol \alpha}}{(2\pi i)^N} \exp(- \alpha ^\dagger  \alpha + u^\dagger_{d'} \alpha + \alpha^\dagger v_d)
\end{equation}
where $\alpha = (\alpha_1, \cdots , \alpha_N)$, $M = e^{-(\beta + it)\boldsymbol \omega}$, $u_d = (\Theta_{1-d}, \cdots , \Theta_{N-d})$, and $v_d = M^\dagger u_d$. Shifting the variables of integration, this can then be simplified to 
\begin{equation}
    I(u^\dagger_{d'}, v_d ):=e^{u^\dagger_{d'}v_d}\int \frac{d \boldsymbol{\alpha}' d\bar{\boldsymbol \alpha}'}{(2\pi i)^N} e^{-\alpha'^\dagger \alpha'} = e^{u^\dagger_{d'}v_d}
\end{equation}
using the fact that the latter half of the expression is equivalent to identity. As done previously, the generating function $I(u^\dagger_{d'}, v_d)$ may then be used to compute matrix elements involving strings of bosonic operators inserted between the coherent state factors in \cref{eq: ft_int}. Insertion of $b_z$ between the rightmost coherent state factors introduces a linear factor of $\alpha_z$ into the coherent state integral, and can be interpreted as a derivative of the generating function with respect to $u^\dagger_{d'}$
\begin{equation}
    \int \frac{d\boldsymbol \alpha d \bar{\boldsymbol \alpha}}{(2\pi i)^N} \cdots \bra{\boldsymbol \Theta_{-d'}}b_z \ket{\boldsymbol \alpha} \rightarrow e^T_{z}\cdot \frac{\partial }{\partial u^\dagger_{d'}}I(u^\dagger_{{d'}}, v_d) = e^T_z \cdot v_de^{u^\dagger_{d'}v_d}.
\end{equation}
Similarly, insertion of $b^\dagger_z$ between the leftmost coherent state factors in \cref{eq: ft_int} introduces a linear factor of $[\alpha^\dagger M^\dagger]_z$ which may be interpreted as a derivative with respect to $v_d$ as 
\begin{equation}
     \int \frac{d\boldsymbol \alpha d \bar{\boldsymbol \alpha}}{(2\pi i)^N} e^{-\sum_z \bar{\alpha}_z \alpha_z} \bra{\boldsymbol \sigma (\beta, t)}b^\dagger_z \ket{\boldsymbol \Theta_{-d}}\cdots \rightarrow e^T_z \cdot (M^\dagger)^T \frac{\partial}{\partial v_d}I(u^\dagger_{d'}, v_d)= e^T_z \cdot (M^* u^*_{d'})e^{u^\dagger_{d'}v_d}.
\end{equation}
Matrix elements involving a greater number of bosonic operator insertions may be computed by evaluating the corresponding chained higher-derivatives detailed above.
\section{Analysis of the optical conductivity}\label{app:oc_pert}
\subsection{Perturbative analysis}
In this appendix, we develop an intuition for the qualitative differences which arise in the observed 2D Holstein and BSSH optical conductivities. As the coupling we use ($g=1$) is comparatively weak in the two-dimensional case, we utilize one-phonon perturbation theory on top of the non-interacting electron--phonon problem with ground state at zero-momentum $\ket{g^{(0)}} = c^\dagger_0\ket{\text{vac}}$. For simplicity, we will assume a single electronic band, as in the Holstein and BSSH models.

Defining the unperturbed one-phonon states with zero total momentum,
\begin{equation}
    \ket{\mu_\mathbf{q}^{(0)}}: = c^\dagger_{\mathbf{q}} b^\dagger_{\mu_{-\mathbf{q}}}\ket{\text{vac}},
\end{equation}
we have that these states have, for the unperturbed Hamiltonian 
\begin{equation}
    \hat{H}_0 = \sum_\mathbf{k} \epsilon_\mathbf{k} c^\dagger_\mathbf{k} c_\mathbf{k} + \sum_{\mu_\mathbf{q}}\omega_{\mu_\mathbf{q}} b^\dagger_{\mu_\mathbf{q}} b_{\mu_\mathbf{q}},
\end{equation}
an unperturbed energy of $E^{(0)}_{\mu_\mathbf{q}} = \epsilon_\mathbf{q} + \omega_{\mu_{-\mathbf{q}}}$. For the 2D Holstein and BSSH models, inserting the tight-binding dispersion and single optical phonon frequency $\omega_0$ allows us to simplify this to 
\begin{equation}
    E^{(0)}_{\mu_{\mathbf{q}}} = \omega_0 -2t(\cos q_x + \cos q_y).
\end{equation}
Using the bare-electronic energy of $E^{(0)}_0=-4t$, we may then define the energy difference
\begin{equation}
\begin{aligned}
    \Delta_{\mathbf{q}} &= E^{(0)}_{\mu_\mathbf{q}} - E^{(0)}_0 \\
    &= \omega_0 + 2t(2 - \cos q_x -\cos q_y).
\end{aligned}
\end{equation}
This equation has the simple interpretation as the excitation energy required to create a phonon and move the electron from the band minimum to a finite momentum. From this, we see that the one-phonon continuum energetically extends from $\omega_0$ at $\mathbf q = (0,0)$ to $\omega_0 + 8t$ at the zone edge $\mathbf q =(\pi, \pi)$.

For a generic electron--phonon interaction term $\hat{V}$, we then have from first-order perturbation theory \cite{Sakurai_Napolitano_2020} the interacting eigenstates to first-order in the coupling, defining $V_{\mu_\mathbf{q}} = \langle{\mu_\mathbf{q}^{(0)}}|\hat{V}|{g^{(0)}}\rangle$,
\begin{equation}
    |g^{(1)}\rangle = |{g^{(0)}}\rangle - \sum_{\mu_\mathbf{q}}\frac{V_{\mu_\mathbf{q}}}{\Delta_\mathbf{q}} |\mu_\mathbf{q}^{(0)}\rangle 
\end{equation}
and
\begin{equation}
    |\mu_\mathbf{q}^{(1)}\rangle = |\mu_\mathbf{q}^{(0)}\rangle + \frac{V^*_{\mu_\mathbf{q}}}{\Delta_\mathbf{q}} |{g^{(0)}}\rangle.
\end{equation}
We further decompose the relevant current operator $\hat{j}_x$ exactly into a component independent of and linear in the coupling,
\begin{equation}
\label{eq:current_decomp}
    \hat{j}_x = \hat{j}^{(0)}_x + \hat{j}_x ^{(1)},
\end{equation}
after which the matrix elements in the regular part of the optical conductivity \cref{eq:reg_oc} can be written
\begin{equation}
\label{eq:first_order_mat}
\begin{aligned}
    \mathcal{M}_{\mu_\mathbf{q}} &= \langle \mu_\mathbf{q}^{(1)}|\hat{j}_x |g^{(1)}\rangle \\
    &\approx -\sum_{\mu_\mathbf{q}'}\frac{V_{\mu_\mathbf{q}'}}{\Delta_\mathbf{q}}\langle{\mu^{(0)}_\mathbf{q}}| \hat{j}^{(0)}_x|{\mu '}^{(0)}_\mathbf{q} \rangle + \frac{V^*_{\mu_\mathbf{q}}}{\Delta_\mathbf{q}}\langle g^{(0)}| \hat{j}^{(0)}_x|g^{(0)}\rangle  + \langle \mu_\mathbf{q}^{(0)}| \hat{j}^{(1)}_x | g^{(0)}\rangle 
\end{aligned}
\end{equation}
to first order. For the 2D nearest-neighbor tight-binding dispersions we consider, $\hat{j}_x^{(0)}$ takes the momentum space form
\begin{equation}
    \hat{j}_x^{(0)} = 2t\sum_{\mathbf{k}}\sin k_x c^\dagger_{\mathbf{k}}c_\mathbf{k}.
\end{equation}
The BSSH model additionally has a phonon-mediated current term of the form
\begin{equation}
    \hat{j}^{(1)}_x = -\frac{2g}{\sqrt{N}}\sum_{\mathbf{k}\mathbf{q}}e^{iq_x/2} \sin(k_x + \frac{q_x}{2}) c^\dagger_{\mathbf{k} + \mathbf q}c_\mathbf{k} (b^\dagger_{x_\mathbf{-q}} + b_{x_\mathbf{q}})
\end{equation}
for the $x$-direction. Now, letting $v_x(\mathbf k) = \frac{\partial \epsilon_{\mathbf k}}{\partial k_x} = 2t\sin k_x$, we have that the general first-order matrix element (\cref{eq:first_order_mat}) can be written as 
\begin{equation}
    \mathcal{M}_{\mu_\mathbf q} = -\frac{V_{\mu_{\mathbf q}}}{\Delta_\mathbf{q}} v_x(q) + \frac{V^*_{\mu_\mathbf{q}}}{\Delta_{\mathbf q}}v_x(0) + J^{(1)}_{\mu_{\mathbf q}},
\end{equation}
where we have defined $J_{\mu_{\mathbf q}}^{(1)} = \langle \mu^{(0)}_{\mathbf q}|\hat{j}^{(1)}_x | g^{(0)}\rangle$. A further simplification is obtained by noting that the electronic band minimum occurs at $\mathbf{k} = 0$, which results in the second term containing $v_x(0)$ vanishing. We now substitute the current operators for Holstein and BSSH into these general expressions.

For the Holstein model, the first-order coupling element may be simply evaluated to be 
\begin{equation}
    V_{\mathbf q}^{\text{H}} = \frac{g}{\sqrt{N}}.
\end{equation}
As the Holstein model supports a current derived exactly from the bare electronic operator, with no component coming from the coupling, the relevant optical conductivity matrix element then becomes
\begin{equation}
    \mathcal{M}^\text{H}_{\mathbf q} = - \frac{2gt}{\sqrt{N} \Delta_{\mathbf q} }\sin(q_x).
\end{equation}
At the one-phonon level, the regular part of the optical conductivity then becomes
\begin{equation}
    \sigma ^{\text{H}}_\text{reg} (\omega) = \frac{\pi}{V}\sum_{\mathbf q}\frac{4t^2g^2}{N \Delta_q^3} \sin^2(q_x)\delta(\omega - \Delta_{\mathbf{q}}).
\end{equation}

On the other hand, the BSSH model has a current piece contributed by the electron--phonon coupling and thus has a non-zero $J^{(1)}_{\mu_{\mathbf q}}$. Evaluating this expression yields
\begin{equation}
    \begin{aligned}
        J^{(1)}_{\mu_\mathbf q} &= -\frac{2g}{\sqrt{N}}\sum_{\mathbf{k}' \mathbf{q}'}e^{iq_x/2}\sin(k_x' + \frac{q_x'}{2})\langle \mu^{(0)}_\mathbf{q}| c^\dagger_{\mathbf{k}' + \mathbf{q}'}c_\mathbf{k} (b^\dagger_{x_{-\mathbf{q}'}} + b_{x_{\mathbf{q}'}})| g^{(0)}\rangle\\
        &= -\frac{2g}{\sqrt{N}} e^{iq_x/2}\sin(\frac{q_x}{2}) \delta_{\mu x}.
    \end{aligned}
\end{equation}
Additionally, the first piece contributes as 
\begin{equation}
\begin{aligned}
    V^{\text{BSSH}}_{\mu_{\mathbf q}} &= \frac{g}{\sqrt{N}}\sum_{\mathbf k' \mu'_{\mathbf q'}}[e^{(ik'_{\mu '} + q'_{\mu'})} + e^{-ik'_{\mu '}}] \langle{\mu^{(0)}_\mathbf q}|c^\dagger_{\mathbf{k}' + \mathbf{q}'}c_{\mathbf{k}'} (b^\dagger_{\mu'_{-\mathbf q '}} + b_{\mu'_{\mathbf{q}'}})| g^{(0)}\rangle \\
    &= \frac{g}{\sqrt{N}}(e^{iq_\mu} + 1)\\
    &=\frac{2g}{\sqrt{N}}e^{iq_\mu/2}\cos(\frac{q_\mu}{2}).
\end{aligned}
\end{equation}
The relevant BSSH optical matrix element then becomes
\begin{equation}
    \mathcal{M}^\text{BSSH}_{\text{reg}} = - \frac{4gt}{\sqrt{N} \Delta_\mathbf q}e^{iq_\mu/2}\cos(\frac{q_\mu}{2})\sin(q_x) -\frac{2g}{\sqrt{N}} e^{iq_x/2}\sin(\frac{q_x}{2})\delta_{\mu x}.
\end{equation}
Squaring this matrix element then yields the BSSH optical conductivity 
\begin{equation}
\sigma^\text{BSSH}_\text{reg}(\omega) = \frac{4\pi g^2}{VN}\sum_{\mathbf{q}} \bigg[\frac{\sin^2(\frac{q_x}{2})}{\Delta_{\mathbf{q}}}\Big[1 +\frac{4t\cos^2(\frac{q_x}{2})}{\Delta_\mathbf{q}}\Big]^2 + \frac{4t^2\sin^2(q_x)\cos^2(\frac{q_y}{2})}{\Delta_\mathbf{q}^3}\bigg]\delta(\omega - \Delta_{\mathbf{q}}),
\end{equation}
where the last term is an indirect Holstein-like contribution.
\subsection{Sector decomposition}
\begin{figure}[hb]
    \centering
    \includegraphics[width=0.4\linewidth]{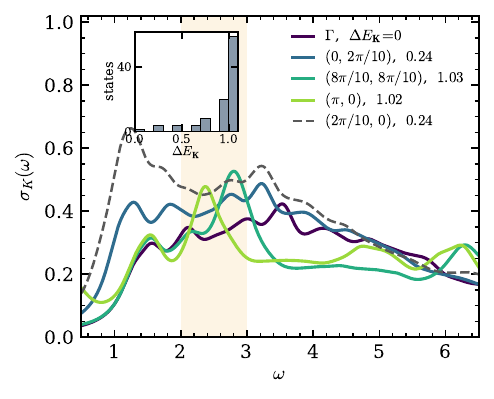}
    \caption{\textbf{Sector-resolved BSSH optical response.} Regular optical conductivity computed from dCC-1-S1-(1,1) for representative total-momentum sectors on a $10\times10$ lattice $(g=t=\omega = 1)$. Energy differences from the band-minimum are given in the legend $\Delta E_\mathbf{K} = E_\mathbf{K} - E_\Gamma$. The shaded band marks the region of the finite-temperature absorption peak. A Lorentzian broadening of $\eta = 0.25$ is used to emphasize the qualitative envelopes of the different absorption profiles.}
    \label{fig:2d_oc_sector}
\end{figure}
The analysis above describes absorption out of the ground state. At finite temperature, initial states of nonzero total crystal momentum additionally contribute, and we characterize their absorption profiles with our tangent space response framework in \cref{fig:2d_oc_sector}. The inset shows the distribution of sector energies relative to the band minimum. The large number of states near the band maximum (capped by hybridization with the one-phonon continuum) and the large absorption feature of such states in the $\omega \simeq 2-3$ region underlie the interpretation of the dominant finite-temperature absorption feature in Section~\ref{subsec:optcond}.
\twocolumngrid
\end{document}